\documentclass[10pt]{article}
\usepackage[a4paper, total={6in, 8in}]{geometry}
\usepackage[usenames,dvipsnames,svgnames,table]{xcolor} 

\usepackage{algorithm}
\usepackage{algorithmic}
\floatname{algorithm}{Algorithm}

\usepackage{longtable}
\usepackage{hhline}
\usepackage{float}
\newfloat{algorithm}{tb}{lop}

\usepackage{enumitem} 
\usepackage{rotfloat} 
\usepackage{graphicx}
\usepackage{epsfig}
\usepackage{amssymb, amsmath, amsthm}
\usepackage{verbatim}
\usepackage{hyperref}
\usepackage{natbib}
\usepackage{authblk}
\usepackage{multirow}
\usepackage{subcaption} 
\usepackage{array}
\newcolumntype{L}{>{\centering\arraybackslash}m{0.1\linewidth}}

\newtheorem{theorem}{Theorem}[section]

\newtheorem{proposition}[theorem]{Proposition}

\newtheorem{definition}{Definition}

\DeclareMathOperator*{\argmin}{argmin}
\DeclareMathOperator*{\argmax}{argmax}

\newcommand{\Exp}{\textrm{Exp}}
\newcommand{\Log}{\textrm{Log}}

\newcommand{\bfu}{\mathbf{u}}
\newcommand{\bfv}{\mathbf{v}}

\newcommand{\bfx}{\mathbf{x}}
\newcommand{\bfy}{\mathbf{y}}
\newcommand{\bfz}{\mathbf{z}}
\newcommand{\bfmu}{\boldsymbol{\mu}}

\newcommand{\bfX}{\mathbf{X}}

\newcommand{\bfZ}{\mathbf{Z}}
\newcommand{\bfTheta}{\mathbf{\Theta}}
\newcommand{\bfGamma}{\mathbf{\Gamma}}

\newcommand{\bbS}{\mathbb{S}}
\newcommand{\bbR}{\mathbb{R}}
\newcommand{\bbX}{\mathbb{X}}
\newcommand{\bbRpos}{\mathbb{R}^+}

\newcommand{\calM}{\mathcal{M}}

\newcommand{\rmE}{\textrm{E}}
\newcommand{\rmP}{\textrm{P}}

\newcommand{\fvMF}{f_{\textrm{vMF}}}
\newcommand{\fSN}{f_{\textrm{SN}}}
\newcommand{\fSL}{f_{\textrm{SL}}}

\newcommand{\mleloc}{\hat{\bfmu}_{\textrm{MLE}}}
\newcommand{\mlescale}{\hat{\sigma}_{\textrm{MLE}}}

\newcommand{\Frechet}{Fr\'{e}chet }

\begin{document}
	
\title{On the spherical Laplace distribution}
\author[1]{Kisung You}
\author[1]{Dennis Shung}
\affil[1]{Department of Internal Medicine, Yale School of Medicine}
\date{}

\maketitle
\begin{abstract}
The von Mises-Fisher (vMF) distribution has long been a mainstay for inference with data on the unit hypersphere in directional statistics. The performance of statistical inference based on the vMF distribution, however, may suffer when there are significant outliers and noise in the data. Based on an analogy of the median as a robust measure of central tendency and its relationship to the Laplace distribution, we proposed the spherical Laplace (SL) distribution, a novel probability measure for modelling directional data. We present a sampling scheme and theoretical results on maximum likelihood estimation. We derive efficient numerical routines for parameter estimation in the absence of closed-form formula. An application of model-based clustering is considered under the finite mixture model framework. Our numerical methods for parameter estimation and clustering are validated using simulated and real data experiments.
\end{abstract}


\section{Introduction}\label{sec1:intro}

A prominent discipline in modern data science is to learn from data equipped with certain geometric constraints. The program often assumes that data reside on Riemannian manifolds \cite{lee_1997_RiemannianManifoldsIntroduction} and has attracted much attention over the last decades in both theoretical foundation \cite{pennec_2006_IntrinsicStatisticsRiemannian, bhattacharya_2012_NonparametricInferenceManifolds, patrangenaru_2016_NonparametricStatisticsManifolds} and practical applications across many fields including  computer vision \cite{bissacco_2001_RecognitionHumanGaits, aggarwal_2004_SystemIdentificationApproach} and medical image analysis  \cite{pennec_2020_RiemannianGeometricStatistics, fletcher_2009_GeometricMedianRiemannian, you_2021_RevisitingRiemannianGeometry}. In statistics, the relevant research area has long been known as directional statistics  \cite{mardia_2000_DirectionalStatistics, ley_2017_ModernDirectionalStatistics} whose objects of interests include directions, axes, and rotations. Especially, directions seem to take the largest portion in the field for its popularity as a number of applications involve observations on the unit hypersphere, which is a model manifold of constant positive curvature. 

An off-the-shelf choice of probability measure on the unit hypersphere is the von Mises-Fisher (vMF) distribution \cite{fisher_1953_DispersionSphere, mardia_2000_DirectionalStatistics}, which is a location-scale family distribution that has higher concentration of mass near a mean direction. Recently, the spherical normal (SN) distribution \cite{hauberg_2018_DirectionalStatisticsSpherical} was proposed as an alternative by adopting the geodesic distance on the unit hypersphere. Both distributions are Gaussian-like laws since they are defined using negated squared distances to their location parameters within an exponential function. For both distributions, it is straightforward to see that the maximum likelihood estimate of a location parameter corresponds to the quantity called \Frechet mean \cite{kendall_1990_ProbabilityConvexityHarmonic}, which is a minimizer of sum of squared distances.

While these distributions may play an important role in statistical inference, their performance may deteriorate when there exist outliers or noise of large magnitude. In classical robust statistics \cite{huber_1981_RobustStatistics}, it is well acknowledged that the median is a robust measure of central tendency. This has been translated in the field of statistics on manifolds to replace \Frechet mean with \Frechet median, which minimizes sum of distances  \cite{fletcher_2009_GeometricMedianRiemannian, arnaudon_2013_MediansMeansRiemannian}. Hence, it is natural to ask what distribution is related to the concept of \Frechet median on the unit hypersphere as the vMF and SN distributions to the \Frechet mean. 

This motivates our paper to propose the spherical Laplace (SL) distribution. As its name entails, the SL distribution generalizes the Laplace distribution on the real line onto the unit hypersphere by defining its density function using  a scaled distance to a mean location rather than squared distance. A leading contribution of this paper is to establish the SL distribution and derive an explicit, simplified form of a normalizing constant amicable for numerical approximation. We also consider a rejection-based sampling scheme to draw random samples from the SL distribution. 

Another class of our contribution is on parameter estimation in the sense of maximum likelihood, which is one of the fundamental tasks for statistical inference. We show that under mild support conditions the SL distribution admits unique parameter estimates. Since the maximum likelihood estimates are  solutions of nonlinear equations, we present algorithms for numerical optimization of the parameters. 

We consider model-based clustering \cite{bouveyron_2019_ModelbasedClusteringClassification} as a major application of the SL distribution. We employ the finite mixture model \cite{mclachlan_2019_FiniteMixtureModels}, which is also a major tool for density estimation, for model-based clustering of data on the unit hypersphere using the SL distribution as its components. Computation for the SL mixture is heavily dependent on a generalized version of the parameter estimation algorithms where observations are given fixed weights. We present computational pipeline in detail for the SL mixture in a similar fashion to its predecessors \cite{banerjee_2005_ClusteringUnitHypersphere, hornik_2014_MovMFPackageFitting}. Computational apparatus for the SL distribution is implemented in an \textsf{R} package called \texttt{Riemann}, which is available on CRAN (\url{https://CRAN.R-project.org/package=Riemann}) and GitHub (\url{https://github.com/kisungyou/Riemann}).

The rest of this paper is organized as follows. We start by introducing rudimentary background on geometry of hypersphere and describing how vMF and SN distributions are related in Section \ref{sec2:background}. Section \ref{sec3:SL} presents the SL distribution along with a rejection sampler and theoretical results regarding maximum likelihood estimation. We present numerical routines for parameter estimation in Section \ref{sec4:parameter}. In Section \ref{sec5:clustering}, the SL mixture is presented in technical details. The distribution and proposed algorithms are validated via simulated and real data experiments in Section \ref{sec6:experiments}. We conclude by discussing potential issues and a list of topics for future studies in Section \ref{sec7:conclusion}.

\section{Background}\label{sec2:background}

We start this section by introducing notations throughout the paper. Let $\bbS^p = \lbrace \bfx \in \bbR^{p+1}: \|\bfx\| = 1 \rbrace$ be a $p$-dimensional unit sphere in a $(p+1)$-dimensional Euclidean space $\bbR^{p+1}$ where $\| \cdot \|$ is the standard $\ell_2$ norm. For a metric space $(\bbX, \delta)$, an open ball of radius $\epsilon > 0$ centered at $\bfx \in \bbX$ is denoted as $B(\bfx, \epsilon) = \lbrace \bfy \in \bbX~|~ \delta(\bfx, \bfy) < \epsilon \rbrace$. For two points $\bfx, \bfy \in \bbS^p$, the distance $d(\bfx, \bfy)$ is reserved to denote the geodesic or shortest-path distance on a sphere. A superscript $^{(t)}$ indicates an object at iteration $t$. The set of positive real numbers is denoted as $\bbRpos$. 

\subsection{Geometry of hypersphere}

The unit hypersphere is a model space for a simply connected manifold of constant positive curvature \cite{carmo_1992_RiemannianGeometry, lee_1997_RiemannianManifoldsIntroduction}. We review basic properties of the unit sphere as a Riemannian manifold and explicit forms of computational operations \cite{absil_2008_OptimizationAlgorithmsMatrix}.

\begin{figure}[ht]
	\centering
	\includegraphics[width=0.4\linewidth]{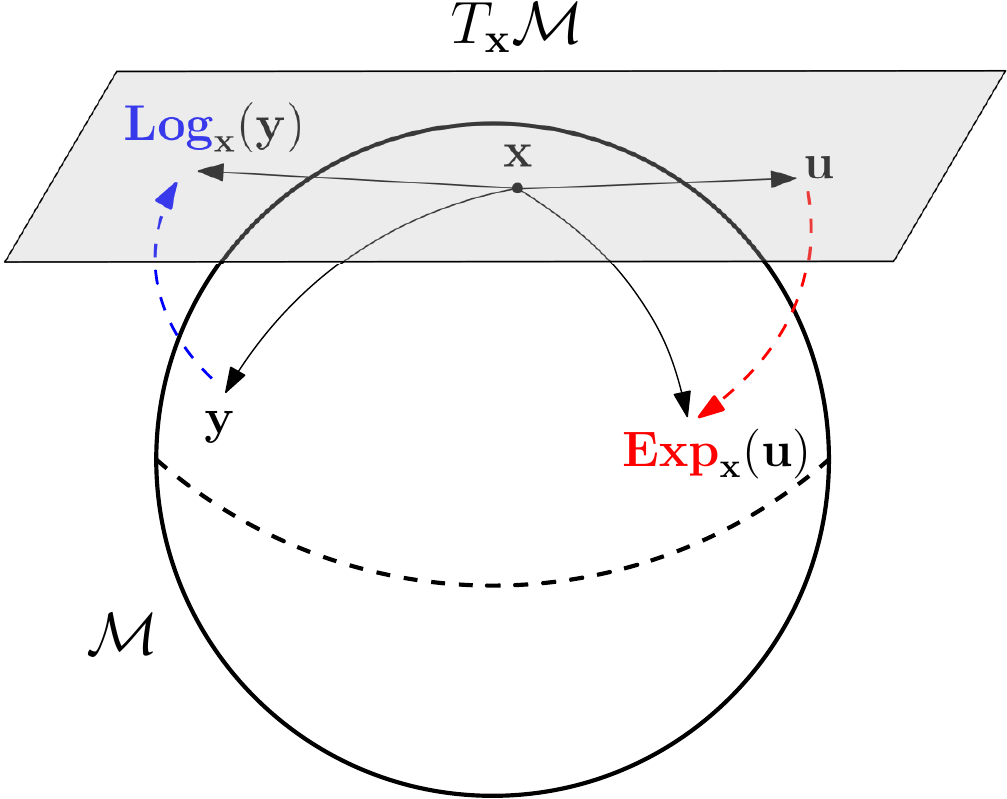}
	\caption{Visual representation of exponential ({\color{red}\bf red}) and logarithmic ({\color{blue} \bf blue}) maps on a Riemannian manifold $\calM$ for some points $\bfx, \bfy \in \calM$ and a tangent vector $\bfu \in T_\bfx \calM$. }
	\label{fig:TwoMaps}
\end{figure}

For some $\bfx \in \bbS^p$, the tangent space is characterized as $T_\bfx \bbS^p = \lbrace \bfz \in \bbR^{p+1}~|~ \langle \bfx, \bfz \rangle = 0 \rbrace $ with a standard inner product $\langle \cdot, \cdot \rangle$.  The Riemannian metric for $\bbS^p$ is a canonical metric, i.e., $g_\bfx (\bfu, \bfv) = \langle \bfu, \bfv\rangle$ for all $\bfu, \bfv \in T_\bfx \bbS^p$. The shortest path connecting any two points $\bfx, \bfy \in \bbS^p$ is along the great circle in that the geodesic distance is given by $d(\bfx, \bfy) = \text{arccos}(\langle \bfx, \bfy \rangle)$. Given a pair $(\bfx, \bfu) \in \calM \times T_\bfx \calM$ for some Riemannian manifold $\calM$, there exists a unique geodesic $\gamma : [0,1] \rightarrow \calM$ such that $\gamma (0) = \bfx$ and $\gamma' (0) = \bfu$. An exponential map is defined by $\Exp_\bfx (\bfu) = \gamma ( 1)$, which maps a vector in some tangent space to the manifold itself whose inverse is called a logarithmic map, as shown in \figurename \ref{fig:TwoMaps}. When $\calM = \bbS^p$, two maps are endowed with explicit formula. Define an operator  $\textrm{Proj}_\bfx (\bfz) = \bfz - \langle \bfx, \bfz \rangle \bfx$ that projects some vector $\bfz \in \bbR^{p+1}$ onto the tangent space $T_\bfx \bbS^p$. For $\bfx, \bfy \in \bbS^p$ and $\bfu \in T_\bfx \bbS^p$, we have the following expressions;
\begin{subequations}
	\begin{align}
	\textrm{(exponential map)}\quad &\quad \text{Exp}_\bfx (\bfu) = \cos (\|\bfu\|) x + \frac{\sin(\|\bfu\|)}{\|\bfu\|} \bfu, \label{eq:map_exponential} \\
	\textrm{(logarithmic map)}\quad &\quad \text{Log}_\bfx (\bfy) = \frac{d(\bfx,\bfy)}{\|\text{Proj}_\bfx (\bfy-\bfx)\|} \text{Proj}_\bfx (\bfy-\bfx). \label{eq:map_logarithmic}
	\end{align}	
\end{subequations}
An exponential map, by its construction, is only locally defined. The injectivity radius $\textrm{inj}(\calM)$ of a manifold $\calM$ is a maximal radius of a geodesic ball where the exponential map is well-defined and a diffeomorphism, which is $\pi$ for the unit hypersphere.

\subsection{The von Mises-Fisher and spherical normal distributions}

On the $p$-dimensional unit hypersphere, the vMF is one of the most popular distributions as an exponential family distribution that has seen many applications  \cite{banerjee_2005_ClusteringUnitHypersphere, bhattacharya_2012_NonparametricBayesClassification, gopal_2014_MisesfisherClusteringModels}. Parametrized by two parameters for  location $\bfmu \in \bbS^p$ and scale $\kappa \in \bbRpos$, the vMF distribution has the following density function of the form $\fvMF (\bfx~|~ \boldsymbol{\mu}, \kappa) \propto \exp (\kappa \bfx^\top \bfmu)$. One observation is that the density is characterized by the square of Euclidean distance,
\begin{equation}\label{eq:vMF_squared}
\exp \left( -\frac{\kappa}{2} \|\bfx - \bfmu\|^2 \right) = \exp \left( -\frac{\kappa}{2} \left\lbrace \bfx^\top \bfx  - 2\bfx^\top \bfmu + \bfmu^\top \bfmu \right\rbrace \right) = C(\kappa) \exp (\kappa \bfx^\top \bfmu),
\end{equation}
where the constant $C(\kappa)$ does not affect the term inside an exponential since $\|\bfx\|=\|\bfmu\|=1$. Therefore, we may roughly consider the vMF distribution as a direct adaptation of the isotropic Gaussian distribution onto the unit hypersphere.

It is natural to question whether the use of standard Euclidean distribution in defining a probability measure on the constrained domain of hypersphere is appropriate. This led to a recent proposal of the SN distribution to replace the Euclidean norm with the geodesic distance. For an isotropic version of the SN distribution, the density function, parametrized by
location $\bfmu \in \bbS^p$ and concentration $\lambda \in \bbRpos$ parameters, is of the following form
\begin{equation}
\fSN (\bfx~|~\bfmu,\lambda) \propto \exp \left(-\frac{\lambda}{2} d^2 (\bfx, \bfmu)\right).
\end{equation}
We close this section by discussing the connection between vMF and SN distributions. As we observed in \eqref{eq:vMF_squared}, both laws are parametrized by scaled square of some distances from a location parameter. In the literature of statistics on manifolds, these two formulations correspond to the dichotomy of \textit{intrinsic} and \textit{extrinsic} frameworks \cite{bhattacharya_2012_NonparametricInferenceManifolds}. Roughly speaking, the intrinsic approach measures dissimilarity of two points on a manifold by the shortest-path geodesic joining the two, while the extrinsic framework uses a standard norm after equivariant embedding of the data on manifold onto the Euclidean space \cite{wasserman_1969_EquivariantDifferentialTopology}. Therefore, vMF and SN distributions may be regarded as extrinsic and intrinsic Gaussian-like probability laws on the sphere, respectively. Hence, this motivates to define a spherical extension of the Laplace distribution by reducing the order of exponent from the SN density function from 2 to 1. 

\section{The spherical Laplace distribution}\label{sec3:SL}
\subsection{Definition}

We propose the spherical Laplace (SL) distribution, which is an isotropic location-scale family distribution on the unit hypersphere $\bbS^p$ of $p\geq 1$. 

\begin{figure}[ht]
	\centering
	\begin{subfigure}[b]{0.3\textwidth}
		\centering
		\includegraphics[width=\textwidth]{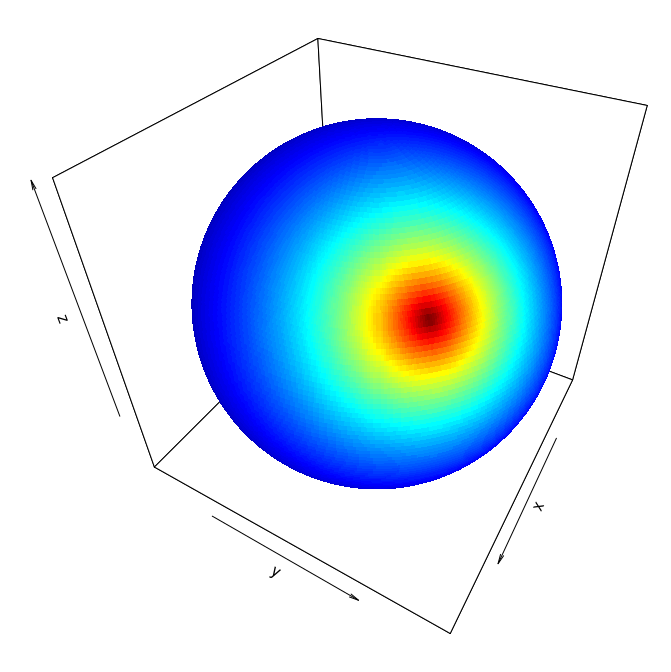}
		\caption{$\sigma=0.5$}
	\end{subfigure}
	\hfill
	\begin{subfigure}[b]{0.3\textwidth}
		\centering
		\includegraphics[width=\textwidth]{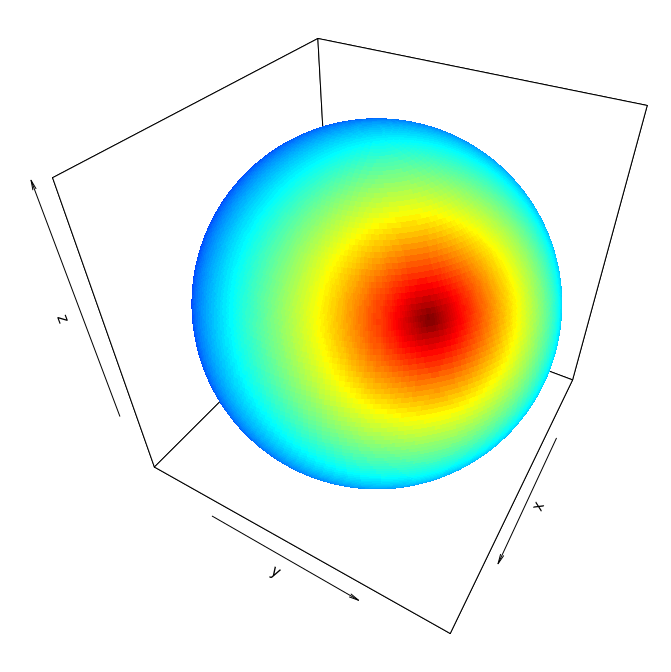}
		\caption{$\sigma=1$}
	\end{subfigure}
	\hfill
	\begin{subfigure}[b]{0.3\textwidth}
		\centering
		\includegraphics[width=\textwidth]{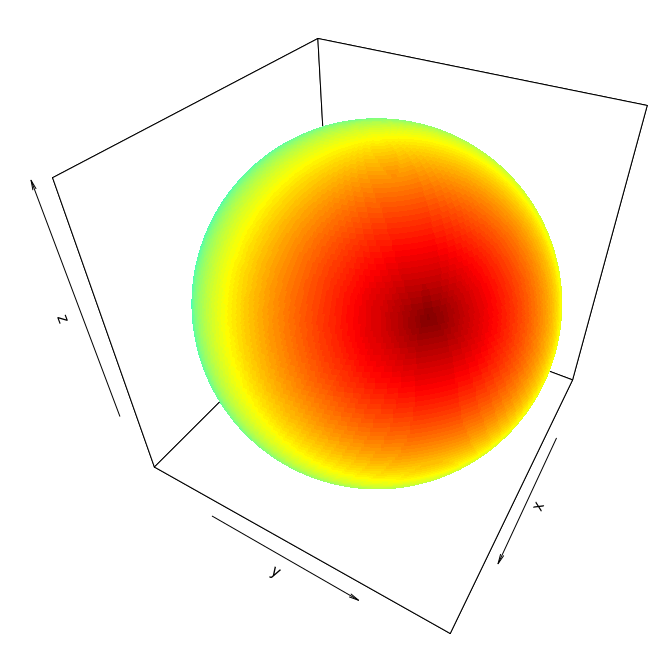}
		\caption{$\sigma=5$}
	\end{subfigure}
	\caption{Densities of the SL distribution on $\bbS^2$ with different values of the scale parameter $\sigma$ with $\bfmu = (1,1,1)/\sqrt{3}$. For each plot, regions that are colored in  {\color{red} \bf red} indicate higher densities compared to the area colored as {\color{blue}\bf blue}.}
	\label{fig:three_densities}
\end{figure}

\begin{definition} Let $\bfmu \in \bbS^p$ and $\sigma \in \bbRpos$ be location and scale parameters, respectively. Density function of the SL distribution is defined by
	\begin{equation*}
	\fSL (\bfx~|~ \boldsymbol{\mu}, \sigma) = \frac{1}{C_p (\bfmu, \sigma)} \exp \left( -\frac{d(\bfx, \bfmu)}{\sigma} \right),
	\end{equation*}
	for a normalizing constant $C_p (\bfmu, \sigma)$,
	\begin{equation}\label{eq:normalizing_constant_full}
	C_p (\bfmu, \sigma) = \int_{\bbS^p} \exp \left( -\frac{d(\bfx, \bfmu)}{\sigma} \right) d\bfx.
	\end{equation}
\end{definition}
The scale parameter quantifies the degree of dispersion as shown in \figurename \ref{fig:three_densities}, where a small $\sigma$ leads to concentrated mass around $\bfmu$ and a large $\sigma$ leads to largely dispersed distribution of the mass.

It is trivial to observe that \eqref{eq:normalizing_constant_full} is well defined. The integrand is a smooth,  bounded function due to the fact that $\sigma$ is strictly positive and $d(\bfx, \bfmu)$ is bounded in $(0,\pi)$. Since the domain is compact, the integral is finite. However, evaluation of \eqref{eq:normalizing_constant_full} involves multidimensional integration over the constrained domain, which makes practical usage of the distribution prohibitive. Although a closed-form formula is not available, we show in the following proposition that the expression for normalizing constant can be transformed into a single-variable integration over the bounded domain. 

\begin{proposition}\label{theory_proposition} The normalizing constant \eqref{eq:normalizing_constant_full} can be written as a univariate integral as
	\begin{equation}\label{eq:normalizing_constant_simplified}
	C_p (\bfmu, \sigma) = A_{p-1} \int_{r=0}^\pi \exp \left( -\frac{r}{\sigma}\right) \sin^{p-1} (r)dr,
	\end{equation}
	where $A_{p-1} = 2\pi^{p/2} / \Gamma(p/2)$ is the hypervolume or surface area of $\bbS^{p-1}$ and $\Gamma(\cdot)$ is the standard Gamma function. 
\end{proposition}
\begin{proof}[\textbf{\upshape Proof:}]
	On a complete Riemannian manifold, one can characterize the geodesic distance in terms of logarithmic maps and Riemannian metric as $d^2(\bfx, \bfmu) = g_{\bfmu} (\Log_{\bfmu} (\bfx), \Log_{\bfmu} (\bfx))$, which is equivalent to $\Log_{\bfmu} (\bfx)^\top \Log_{\bfmu} (\bfx)$ on $\bbS^p$ with a canonical metric. By a change of variable $\bfu = \Log_{\bfmu} (\bfx)$, a corresponding Jacobian determinant $|\mathbf{J}| = (\sin \|\bfu\| / \|\bfu\|)^{p-1}$ is obtained so that
	\begin{eqnarray}
	C_p (\bfmu, \sigma) &=& \int_{\bbS^p}  \exp \left( -\frac{d(\bfx, \bfmu)}{\sigma} \right) d\bfx =  \int_{\bbS^p}  \exp \left( -\frac{\|\bfu\|}{\sigma}\right) d\bfx \nonumber \\  &=& \int_{\|\bfu \| < \pi} \exp \left( -\frac{\|\bfu\|}{\sigma}\right) \left(\frac{\sin \|\bfu \|}{\|\bfu\|}\right)^{p-1} d\bfu. \label{eq:sub_eqnarray}
	\end{eqnarray}
	We can express the above using spherical coordinate system,
	\begin{eqnarray}
	\eqref{eq:sub_eqnarray} &=& \int_{r=0}^\pi \int_{\varphi_1=0}^\pi \cdots \int_{\varphi_{p-2}=0}^\pi \int_{\varphi_{p-1}=0}^{2\pi} \exp \left(-\frac{r}{\sigma}\right) \left(\frac{\sin r}{r}\right)^{p-1} r^{p-1} \sin^{p-2} (\varphi_1)  \cdots \sin (\varphi_{p-2}) dr d\varphi_1 \cdots d\varphi_{p-1} \nonumber \\
	&=& \int_{\varphi_1=0}^\pi \cdots \int_{\varphi_{p-2}=0}^\pi \int_{\varphi_{p-1}=0}^{2\pi}\sin^{p-2} (\varphi_1)  \cdots \sin (\varphi_{p-2})  d\varphi_1 \cdots d\varphi_{p-1}		 \cdot 
	\int_{r=0}^\pi \exp \left(-\frac{r}{\sigma}\right) \sin^{p-1}(r)dr,\label{eq:sub_surface}
	\end{eqnarray}
	where the first term with nested integrals in \eqref{eq:sub_surface} reduces to the hypervolume $A_{p-1}$ of $\bbS^{p-1}$. 
\end{proof}

Proposition \ref{theory_proposition} does not only lower computational complexity to evaluate the normalizing constant, but also implies that it does not depend on the choice of $\bfmu$. In what follows, we will denote the normalizing constant of the SL distribution as $C_p (\sigma)$ to emphasize its independence on a choice of location $\bfmu$.

\subsection{Sampling}

It is often of fundamental interest to draw random samples for a given probability distribution. We start by employing a standard rejection sampling technique for its ease of use and efficiency \cite{bishop_2006_PatternRecognitionMachine}. We take the SN distribution as a  proposal density for the rejection sampler whose sampling strategy was well studied in \cite{hauberg_2018_DirectionalStatisticsSpherical}. For a SL distribution with parameters $(\bfmu, \sigma) \in \bbS^p \times \bbRpos$, setting a SN concentration parameter as $\lambda = 1/\sigma$ with the same location $\bfmu$ reduces to a simple rule of acceptance-rejection, which is summarized in Algorithm \ref{code:algorithm_sampling}. Technical details regarding the sampling scheme is described in Appendix.

\begin{algorithm}[t]
	\caption{Rejection Sampling from the SL distribution}
	\label{code:algorithm_sampling}
	\begin{algorithmic}
		\REQUIRE parameters $\bfmu, \sigma$.
		\ENSURE a random sample $\bfx$.
		\STATE Set $\lambda = 1/\sigma$.
		\REPEAT 
		\STATE $u \sim \text{Uniform}(0,1)$.
		\STATE $\bfy \sim \fSN(\bfy~|~\bfmu, \lambda)$.
		\STATE $r \leftarrow d(\bfmu, \bfy)$.
		\STATE $\tau \leftarrow \exp( (r^2-2r-\pi^2+2\pi)  / 2\sigma)$.
		\UNTIL $u \leq \tau$.
		\STATE  Take $\bfx \leftarrow \bfy$.
	\end{algorithmic}
\end{algorithm}

While the rejection sampler is a convenient choice to sample from the SL distribution, one drawback is that the efficiency deteriorates when $\sigma \approx 0$. Recall that the distance is upper bounded by $\pi$ for any two points on the unit hypersphere. When a draw $\bfy$ from a corresponding SN distribution is given, the acceptance threshold $\tau$ is defined as 
\begin{equation*}
\tau = \exp\left( \frac{d^2(\bfmu,\bfy)-2d(\bfmu,\bfy) - \pi^2+2\pi}{2\sigma}\right) = \exp\left(\frac{(d(\bfmu,\bfy)-1)^2 - (\pi-1)^2}{2\sigma}\right),
\end{equation*}
where the numerator within an exponential function is always smaller than or equal to 0. It is trivial to observe that $\tau \rightarrow 1$ as $\sigma \rightarrow \infty$, implying that the rejection sampler almost always accepts a new sample for larger values of a scale parameter $\sigma$. On the other hand, when $\sigma \rightarrow 0$, the threshold converges to 0 in that the sampler requires a tremendously larger number of repeated sampling from its proposal distribution. Hence, it is suggested to use other Markov chain Monte Carlo methods such as the Metropolis-Hastings (MH) algorithm \cite{metropolis_1953_EquationStateCalculations, hastings_1970_MonteCarloSampling} in the regime of small $\sigma$ values. In our numerical experiments, we used the MH sampler where a sequence of samples is generated from a Gaussian distribution on the tangent space at an iterate.

\subsection{Maximum likelihood estimation}

We now consider the task of maximum likelihood estimation for parameters of the SL distribution. Let $ \bfx_1, \ldots, \bfx_N \in \bbS^p$ be an i.i.d sample from the SL distribution with two parameters,  location $\bfmu$ and scale $\sigma$. The maximum likelihood estimates $(\mleloc, \mlescale) \in \bbS^p \times \bbRpos$ are maximizers of the following log-likelihood function,
\begin{equation}\label{def:log_likelihood}
\begin{split}
L(\bfmu, \sigma) &= \sum_{n=1}^N  \log \left\lbrace \frac{1}{C_p(\sigma)} \exp \left(-\frac{d(\bfx_n,\bfmu)}{\sigma}\right)\right\rbrace\\ 
&= -\frac{1}{\sigma }\sum_{n=1}^N d(\bfx_n, \bfmu) - N \log C_p (\sigma).
\end{split}
\end{equation}
One can observe that \eqref{def:log_likelihood} does not admit closed-form formulae of its maximizers because of the nonlinear terms, engendering the need to employ computational approach for parameter estimation, which will be discussed in the next section.

A closer look at \eqref{def:log_likelihood} allows us to easily recognize that the constrained maximization with respect to $\bfmu$ is independent on the scale parameter since
\begin{equation}\label{eq:mle_location}
\mleloc = \underset{\bfmu \in \bbS^p}{\argmax} -\frac{1}{\sigma}\sum_{n=1}^N d(\bfx_n, \bfmu) = \underset{\bfmu \in \bbS^p}{\argmin} \sum_{n=1}^N d(\bfx_n, \bfmu).
\end{equation}
In the literature of manifold-valued statistics, the objective to minimize sum of distances has been known as the \Frechet or geometric median problem \cite{fletcher_2009_GeometricMedianRiemannian, arnaudon_2013_MediansMeansRiemannian}. Once $\mleloc$ is obtained, solving for $\sigma$ is merely a univariate minimization problem as follows,
\begin{equation}\label{eq:mle_scale}
\mlescale = \underset{\sigma\in\bbRpos}{\argmax} -\frac{1}{\sigma} \sum_{n=1}^N d(\bfx_n, \mleloc) - N \log C_p (\sigma) = \underset{\sigma\in\bbRpos}{\argmin~} \frac{S}{\sigma} + \log C_p(\sigma),
\end{equation}
for a fixed constant $S = \sum_{n=1}^N d(\bfx_n, \mleloc)/N$. We show existence and uniqueness of maximum likelihood estimates in the following theorem under some conditions.

\begin{theorem}\label{theory_theorem}
	Let $\bfx_1, \ldots, \bfx_N$ be an i.i.d sample on a $p$-dimensional unit hypersphere $\bbS^p$. If the sample is contained in an open geodesic ball $B(\bfx, \pi/4)$ for some $\bfx \in \bbS^p$ and not totally contained in any geodesic, maximum likelihood estimates $(\mleloc, \mlescale)$ uniquely exist. 
\end{theorem}
\begin{proof}[\textbf{\upshape Proof:}] We first use the characterization of maximum likelihood estimate for location parameter of the SL distribution as the \Frechet median. Conditions for existence and uniqueness have been extensively studied on a general Riemannian manifold $\calM$ \cite{yang_2010_RiemannianMedianIts, arnaudon_2013_MediansMeansRiemannian}, whose statements are phrased as follows for completeness. Let $\bar{B}(\bfx, \rho)$ and $\Delta$ denote a closed geodesic ball of radius $\rho$ centered at $\bfx \in \calM$ and an upper bound of sectional curvatures in $\bar{B}(\bfx, \rho)$, respectively. Theorem 3.1 of \cite{yang_2010_RiemannianMedianIts} states that if a sample is not totally contained in any geodesic and the radius satisfies
	\begin{equation*}
	\rho < \min \left\lbrace \frac{\pi}{4\sqrt{\Delta}}, \frac{\textrm{inj} (\bar{B}(\bfx, \rho))}{2} \right\rbrace,
	\end{equation*}
	the \Frechet median exists and is unique because the objective function becomes strictly convex under the stated conditions. The unit hypersphere has a constant sectional curvature so that $\Delta = 1$ and an injectivity radius is $\pi$ at all points. Therefore, the maximal convexity radius $\rho$ is $\min \lbrace  \pi/4, \pi/2 \rbrace = \pi/4$ on $\bbS^p$. On top of a support condition of a random sample being not totally contained in any geodesic, this establishes the existence and uniqueness of $\mleloc$. 
	
	We now turn to examine the scale parameter. Let $g(\sigma) = S/\sigma + \log C_p (\sigma)$ with a known constant $S \in [0, \pi]$ as the geodesic distance between any two points on the sphere is bounded above by the injectivity radius $\pi$. It needs to be shown that $g(\sigma)$ admits a unique critical point.	The first-order condition $g'(\sigma) = 0$ is equivalent to whether $-S C_p(\sigma) + \sigma^2 C_p'(\sigma)=0$. Hence, it is sufficient to show if $G(S) = -S C_p(\sigma) + \sigma^2 C_p'(\sigma)$ has a unique zero in $(0,\pi)$ for any choice of $\sigma$. This implies a bijection between $S$ and $\sigma$, establishing the existence and uniqueness for a critical point of $g(\sigma)$ equivalently. 
	
	First, it is trivial that $G(S)$ is a continuous function since every term consists of smooth, bounded functions on a finite interval. Second, we have $G(0) > 0$ as
	\begin{eqnarray*}
		G(0) &=& \sigma^2 C_p'(\sigma) \\
		&=& \sigma^2 \cdot  A_{p-1} \int_{r=0}^\pi \frac{r}{\sigma^2} \exp \left( -\frac{r}{\sigma}\right) \sin^{p-1}(r)dr \\
		&=& A_{p-1} \int_{r=0}^\pi r \exp  \left( -\frac{r}{\sigma}\right) \sin^{p-1}(r)dr,
	\end{eqnarray*}
	where all terms in the integrand are positive on $(0,\pi)$. On the other end, $G(\pi) < 0$ since  
	\begin{eqnarray*}
		G(\pi) &=& -\pi C_p(\sigma) + \sigma^2 C_p'(\sigma)\\
		&=& A_{p-1} \left\lbrace  -\pi \int_{r=0}^\pi \exp \left(-\frac{r}{\sigma}\right)\sin^{p-1}(r)dr + \sigma^2 \int_{r=0}^\pi \frac{r}{\sigma^2}\exp \left( -\frac{r}{\sigma}\right) \sin^{p-1}(r)dr \right\rbrace \\
		&=& A_{p-1} \int_{r=0}^\pi (-\pi + r) \exp \left( -\frac{r}{\sigma}\right) \sin^{p-1}(r)dr ,
	\end{eqnarray*}
	and $-\pi+r < 0$ while the other terms in the integrand are strictly positive except for a measure zero set. Lastly, $G(S)$ is a monotonically decreasing function. Take $S' \in (0,\pi)$ such that $S < S'$, then 
	\begin{eqnarray*}
		G(S')-G(S) &=& -S' C_p (\sigma) + \sigma^2 C_p'(\sigma) - ( -S C_p (\sigma) + \sigma^2 C_p'(\sigma) ) \\
		&=& (S - S') C_p(\sigma) < 0.
	\end{eqnarray*}
	By the intermediate value theorem and monotonicity, $G(S)$ has a unique zero in $(0,\pi)$ and so does $g'(\sigma)$, which completes the proof. 
\end{proof}

\section{Parameter estimation}\label{sec4:parameter}

It was observed in the previous section that maximum likelihood estimation for parameters of the SL distribution can be decomposed into two components in a sequential manner to solve the standard \Frechet median problem for the location parameter $\bfmu \in \bbS^p$ and a univariate optimization problem for the scale parameter $\sigma \in \bbRpos$. Throughout this section, we consider a more general scenario where each $\bfx_n$ is weighted by a non-negative constant $w_n > 0$ that sums to 1, i.e., $\sum_{n=1}^N w_n = 1$, and present adapted algorithms to the context. This is because it not only reduces to the maximum likelihood estimation when all weights are set to be equal, but also this formulation provides useful apparatus for model-based clustering in the following section. 

\subsection{Estimation of the location}

Given a random sample $\bfx_1, \ldots, \bfx_N \in \bbS^p$ and corresponding non-negative weights $w_1, \ldots, w_N$, the weighted \Frechet median problem is written as 
\begin{equation}\label{eq:objective_location}
\underset{\bfmu\in\bbS^p}{\min~} F(\bfmu) = \underset{\bfmu\in\bbS^p}{\min~} \sum_{n=1}^N w_n d(\bfx_n, \bfmu),
\end{equation}
where $\mleloc$ is a solution to the special case of \eqref{eq:objective_location} if all weights $w_n$'s are equal such as $w_1 = \cdots = w_N = 1/N$. 

We employ a geometric variant of the Weiszfeld algorithm \cite{fletcher_2009_GeometricMedianRiemannian}, whose standard version has long been known in the Euclidean regime \cite{weiszfeld_1937_PointPourLequel}. In general, the Weiszfeld algorithm follows a descent path, making it to belong to a class of gradient descent methods. At each iteration, an iterate is updated by varying amount of step size determined by a fixed rule. This removes potentially large computational burden to tune step sizes by line search at each iteration.

Now a geometric Weiszfeld algorithm is described for the weighted \Frechet median problem, which is summarized in Algorithm \ref{code:algorithm_location}. Roughly speaking,  the gradient $\textrm{grad} F\vert_{\bfx}$ of a scalar-valued function $F$ on a manifold $\calM$ is defined as a vector field, meaning that it takes a value as a tangent vector in $T_{\bfx} \calM$. Once an update is done in the gradient direction, an additional step is necessary   to push an iterate onto the manifold itself \cite{absil_2008_OptimizationAlgorithmsMatrix}. Using the variation of energy, it can be readily shown that $\text{grad}~ d(\bfmu, \bfx)\vert_{\bfmu} = -\Log_{\bfmu} (\bfx) / d(\bfmu, \bfx)$, leading to define a gradient of \eqref{eq:objective_location} at iteration $t$ as
\begin{equation*}
\textrm{grad}~ F\vert_{\bfmu^{(t)}} = -\sum_{n=1}^N \frac{w_n}{d(\bfx_n, \bfmu^{(t)})} \Log_{\bfmu^{(t)}} (\bfx_n) =- \sum_{n=1}^N w_n^{(t)} \Log_{\bfmu^{(t)}} (\bfx_n).
\end{equation*}
The geometric Weiszfeld algorithm updates an iterate by 
\begin{eqnarray}
\bfmu^{(t+1)} &=& \Exp_{\bfmu^{(t)}} \left( -\alpha^{(t)}  \textrm{grad}~ F\vert_{\bfmu^{(t)}} \right) \nonumber \\ 
&=& \Exp_{\bfmu^{(t)}} \left(
\frac{\sum_{n=1}^N w_n^{(t)} \Log_{\bfmu^{(t)}}(\bfx_n)}{\sum_{n=1}^N w_n^{(t)}}
\right),\label{eq:update_location_weiszfeld}
\end{eqnarray}
with an automatically-determined step size $\alpha^{(t)}=1/\sum_{n=1}^N w_n^{(t)}$. In our context, exponential and logarithmic maps are computed using the formula introduced in \eqref{eq:map_exponential} and \eqref{eq:map_logarithmic}.

\begin{algorithm}[t]
	\caption{Weighted \Frechet median computation}
	\label{code:algorithm_location}
	\begin{algorithmic}
		\REQUIRE a random sample $\bfx_1, \ldots, \bfx_N \in \bbS^p$,  weights $w_1, \ldots, w_N$, stopping criterion $\epsilon$.
		\ENSURE $\hat{\bfmu} = \argmin F(\bfmu)$ where $F(\bfmu) =  \sum_{n=1}^N w_n  d (\bfx_n, \bfmu)$ for $\bfmu \in \bbS^p$. 
		\STATE Initialize $\bfmu^{(0)} = \sum_{n=1}^N w_n \bfx_n / \| \sum_{n=1}^N w_n \bfx_n \|$.
		\REPEAT 
		\STATE Compute scaled weights $w_n^{(t)} = w_n / d(\bfmu^{(t)}, \bfx_n)$ for $n\in\lbrace 1,\ldots, N\rbrace$. 
		\STATE Update an iterate $\bfmu^{(t+1)}$ by \eqref{eq:update_location_weiszfeld}.
		\IF{$\bfmu^{(t+1)}$ equals to one of $\bfx_1, \ldots, \bfx_N$}
		\STATE stop the algorithm.
		\ENDIF
		\UNTIL $d(\bfmu^{(t)}, \bfmu^{(t+1)})$ or  $\| \bfmu^{(t)} - \bfmu^{(t+1)} \|$ is smaller than $\epsilon$.
		\STATE  Take $\hat{\bfmu} = \bfmu^{(t+1)}$.
	\end{algorithmic}
\end{algorithm}

We close by discussing some technical components of Algorithm \ref{code:algorithm_location}. At the initial stage, a starting point is taken as $\bfmu^{(0)} = \sum_{n=1}^N w_n \bfx_n / \| \sum_{n=1}^N w_n \bfx_n\|$. This quantity is equivalent to an extrinsic mean of a random sample \cite{bhattacharya_2012_NonparametricInferenceManifolds}, which is also a maximum likelihood estimate of a location parameter from the vMF distribution. When a random sample is adequately concentrated as stated in Theorem \ref{theory_theorem}, an extrinsic mean can be a reasonable choice as it is contained in a convex geodesic ball. Moreover, it only requires elementary arithmetic operations without any need for expensive computational routines to be performed. Second, it is standard to halt iterations when an iterate belongs to a set of given observations. A direct reason is that a scaled weight cannot be evaluated if $d(\bfmu^{(t+1)}, \bfx_n)=0$ for any $n$ at the next iteration. Ad hoc remedies include to re-initialize an algorithm, remove a matching point from computation in the specific step, or pad 0 with a very small positive number, all of which we do not consider in this paper. Lastly, the stopping criterion by sufficiently small incremental geodesic change $d(\bfmu^{(t)}, \bfmu^{(t+1)})$ is based from a fact that the unit hypersphere is a complete Riemannian manifold, which guarantees to measure convergence of a sequence in the sense of Cauchy. One may also choose to use the $\ell_2$ norm-based stopping rule since intrinsic and extrinsic distances converge as two points get sufficiently close on a complete manifold embedded in some Euclidean space. 
\subsection{Estimation of the scale}

Maximum likelihood estimation for the scale parameter $\sigma$ of the SL distribution is merely a univariate minimization of the following objective function
\begin{equation*}
g(\sigma) = \frac{S}{\sigma} + \log C_p (\sigma),
\end{equation*} 
given a fixed constant $S = \sum_{n=1}^N d (\bfx_n, \mleloc) / N$. According to the Theorem \ref{theory_theorem}, the cost function admits a unique critical point. Hence, it can be cast as a root-finding problem of $g'(\sigma) = 0$. 

In order to estimate the scale parameter, we use the Newton-Raphson method \cite{ypma_1995_HistoricalDevelopmentNewton, burden_2016_NumericalAnalysis} for root-finding. When a univariate  function of interest $f:\bbR \rightarrow \bbR$ has continuous derivatives up to order 0, the Newton-Raphson algorithm solves $f(x) = 0$ by updating an iterate as $x^{(t+1)} = x^{{(t)}} - {f(x^{(t)})}/{f'(x^{(t)})}$ with an initial point $x^{(0)}$. By setting $f = g'$, we can easily derive the updating rule to solve $g'(\sigma) = 0$ by  $\sigma^{(t+1)} = \sigma^{(t)} - {g'(\sigma^{(t)})}/{g''(\sigma^{(t)})}$. Define the quantities
\begin{eqnarray*}
	I_0 (\sigma) &=& \int_{r=0}^\pi \exp \left( -\frac{r}{\sigma}\right) \sin^{p-1}(r)dr, \\
	I_1 (\sigma) &=& \int_{r=0}^\pi \frac{r}{\sigma^2} \exp \left( -\frac{r}{\sigma}\right) \sin^{p-1}(r)dr, \\
	I_2 (\sigma) &=& \int_{r=0}^\pi \left(\frac{r^2}{\sigma^4} - \frac{2r}{\sigma^3}\right) \exp \left( -\frac{r}{\sigma}\right) \sin^{p-1}(r)dr,
\end{eqnarray*}
which are integrals related to the normalizing constant and its derivatives. By re-writing derivatives of $g$ with respect to the above integrals, we can simplify the Newton-Raphson update at iteration $t$ as
\begin{eqnarray}
\sigma^{(t+1)} = \sigma^{(t)} - 
\left(
-\frac{S}{(\sigma^{(t)})^2} + \frac{I_1 (\sigma^{(t)})}{I_0 (\sigma^{(t)})} 
\right) \Bigg/ \left(
\frac{2S}{(\sigma^{(t)})^3} + \frac{I_0 (\sigma^{(t)}) I_2 (\sigma^{(t)}) - (I_1 (\sigma^{(t)}))^2}{(I_0 (\sigma^{(t)}))^2}
\right).\label{eq:update_scale_newton_exact}
\end{eqnarray}

When $\sigma$ is close to zero, evaluation of $I_1(\sigma)$ and $I_2 (\sigma)$ may suffer from numerical instability due to presence of $\sigma$ in rational terms, while $I_0 (\sigma)$ is more robust to obtain from a computational point of view. As a result, an alternative to \eqref{eq:update_scale_newton_exact} is considered by computing derivatives with the centered finite difference schemes as follows,
\begin{eqnarray*}
	g'(\sigma) &\approx& \frac{g(\sigma+h) - g(\sigma-h)}{2h}, \\
	g''(\sigma) &\approx& \frac{g(\sigma+h) - 2g(\sigma) + g(\sigma-h)}{h^2},
\end{eqnarray*}
for a sufficiently small step-size $h>0$. By plugging in the approximate derivatives, we get an updating rule
\begin{eqnarray}
\sigma^{(t+1)} = \sigma^{(t)} - \frac{h}{2} \cdot \frac{g(\sigma^{(t)}+h) - g(\sigma^{(t)}-h)}{g(\sigma^{(t)}+h) - 2g(\sigma^{(t)}) + g(\sigma^{(t)}-h)}
.\label{eq:update_scale_newton_approx}	
\end{eqnarray}

We note that an approximate updating rule \eqref{eq:update_scale_newton_approx} has equivalent computational complexity to that of \eqref{eq:update_scale_newton_exact} as both integrate one-dimensional smooth functions three times per  iteration.

\begin{algorithm}[t]
	\caption{Maximum likelihood estimation of the scale parameter}
	\label{code:algorithm_scale}
	\begin{algorithmic}
		\REQUIRE a random sample $\bfx_1, \ldots, \bfx_N \in \bbS^p$, a constant $S$, stopping criterion $\epsilon$.
		\ENSURE $\mlescale = \argmin g(\sigma)$ where $g(\sigma) = S/\sigma + \log C_p (\sigma)$ for $\sigma \in \bbRpos$. 
		\STATE Initialize $\sigma^{(0)}$. 
		\REPEAT 
		\STATE Update an iterate $\sigma^{(t+1)}$ by either \eqref{eq:update_scale_newton_exact} or \eqref{eq:update_scale_newton_approx}.
		\UNTIL $|\sigma^{(t)} - \sigma^{(t+1)}|<\epsilon$.
		\STATE  Take $\mlescale = \sigma^{(t+1)}$.
	\end{algorithmic}
\end{algorithm}

\section{Application to clustering}\label{sec5:clustering}

One of the direct applications for parameter estimation of the SL distribution is probabilistic clustering of sphere-valued data using finite mixture models \cite{mclachlan_2019_FiniteMixtureModels}. The density of a finite mixture model of $K$ SL components is 
\begin{eqnarray*}
	h(\bfx | \mathbf{\Theta}) = \sum_{k=1}^K \pi_k \fSL(\bfx~|~\bfmu_k, \sigma_k),
\end{eqnarray*}
for component weights $\pi_k,~k\in\lbrace 1,\ldots, K\rbrace$ such that $\sum_{k=1}^K \pi_k = 1$ and parameters $\bfTheta = \lbrace \pi_k, \bfmu_k, \sigma_k\rbrace_{k=1}^K$. We follow a standard approach to maximize log-likelihood given a random sample $\bfX = \lbrace \bfx_1, \ldots, \bfx_N\rbrace \subset \bbS^p$ by introducing latent variables for class membership and applying  Expectation-Maximization (EM) algorithm \cite{dempster_1977_MaximumLikelihoodIncomplete}. We refer interested readers to \cite{bishop_2006_PatternRecognitionMachine} for thorough description of the technique. To briefly introduce, denote a binary matrix of latent class memberships as $\bfZ \in \lbrace 0, 1\rbrace^{N\times K}$ such that every row of $\bfZ$ contains only one non-zero element and the joint distribution of $\bfX$ and $\bfZ$ as
\begin{eqnarray*}
	\rmP(\bfX,\bfZ~|~\bfTheta) = \prod_{n=1}^N \prod_{k=1}^K \left( \pi_k \fSL(\bfx_n~|~\bfmu_k, \sigma_k)\right)^{z_{nk}}.
\end{eqnarray*}
The EM algorithm alternates E- and M-steps. At iteration $t$, the E-step evaluates posterior distribution of the latent variable  $\rmP(\bfZ~|~\bfX,\bfTheta^{(t)})$ to compute the complete-data log-likelihood
\begin{eqnarray*}
	Q(\bfTheta~|~\bfTheta^{(t)}) = \sum_{\bfZ} \rmP(\bfZ~|~\bfX, \bfTheta^{(t)}) \log \rmP (\bfX, \bfZ~|~\bfTheta) = \rmE_{\bfZ|\bfX, \bfTheta^{(t)}} \lbrack \log \rmP(\bfX, \bfZ~|~\bfTheta) \rbrack,
\end{eqnarray*}
and the M-step updates all parameters $\bfTheta^{(t+1)}$ by maximizing $Q(\bfTheta~|~\bfTheta^{(t)})$. 

We now elaborate on explicit expressions for updating parameters in a finite mixture of SL distributions under the standard EM framework at iteration $t$. The E-step starts by evaluating posterior of the latent variables, which first reduces to evaluating $\rmE[z_{nk}]$ as follows,
\begin{eqnarray}
\gamma_{nk} := \rmE[z_{nk}] = \frac{\pi_k^{(t)} \fSL(\bfx_n~|~\bfmu_k^{(t)}, \sigma_k^{(t)})}{\sum_{j=1}^K \pi_j^{(t)} \fSL(\bfx_n~|~\bfmu_j^{(t)}, \sigma_j^{(t)})},
\end{eqnarray}
for $(n,k) \in \lbrace 1,\ldots, N\rbrace \times \lbrace 1,\ldots, K\rbrace$. Let $\bfGamma := \gamma_{nk} \in [0,1]^{N\times K}$ denote a matrix of posterior for the latent variables known as soft clustering or membership matrix. An $(n,k)$-th entry of $\bfGamma$ encodes information on how likely an observation $\bfx_n$ belongs to the $k$-th cluster. Given the evaluations, the complete-data log-likelihood can be written as
\begin{eqnarray*}
	Q(\bfTheta~|~\bfTheta^{(t)}) = \sum_{n=1}^N \sum_{k=1}^K \gamma_{nk} \left\lbrace 
	\log \pi_k + \log \fSL (\bfx_n~|~\bfmu_k, \sigma_k)
	\right\rbrace,
\end{eqnarray*}
which is an objective function for maximization in the M-step. For each component $k\in\lbrace 1,\ldots,K\rbrace$,  weight and location parameters are updated as follows,
\begin{eqnarray}
\pi_k^{(t+1)} &=& \frac{\sum_{n=1}^N \gamma_{nk}}{N},\nonumber\\
\bfmu_k^{(t+1)} &=& \underset{\bfmu \in \bbS^p}{\argmin~} \sum_{n=1}^N \gamma_{nk} \cdot d(\bfx_n, \bfmu), \label{eq:EM_location}
\end{eqnarray}
where \eqref{eq:EM_location} is a weighted \Frechet median problem that can be solved using Algorithm \ref{code:algorithm_location} with $w_i = \gamma_{ik},~i\in\lbrace 1,\ldots, N\rbrace$. Lastly, scale parameters are solutions of the following problems,
\begin{eqnarray}\label{eq:EM_scale_hetero}
\sigma_k^{(t+1)} = \underset{\sigma \in \bbRpos}{\argmin~} \left(\frac{\sum_{n=1}^N \gamma_{nk} \cdot d(\bfx_n, \bfmu_k^{(t+1)})}{\sum_{n=1}^N \gamma_{nk}}\right) \cdot \frac{1}{\sigma} + \log C_p (\sigma),
\end{eqnarray}
which can be solved by Algorithm \ref{code:algorithm_scale}.

We now turn to discuss some aspects of the model and EM algorithm. First, one way to regularize varying scales across multiple components is to use a common scale parameter rather than component-specific values, i.e., $\sigma = \sigma_1 = \cdots = \sigma_K$. In this setting, the model is called homogeneous and a common scale parameter is updated by
\begin{eqnarray}\label{eq:EM_scale_homo}
\sigma^{(t+1)} = \underset{\sigma \in \bbRpos}{\argmin~} \left(\frac{\sum_{n=1}^N \sum_{k=1}^K \gamma_{nk} \cdot d(\bfx_n, \bfmu_k^{(t+1)})}{\sum_{n=1}^N \sum_{k=1}^K \gamma_{nk}}\right) \cdot \frac{1}{\sigma} + \log C_p (\sigma).
\end{eqnarray}

Second, computational cost of the algorithm becomes prohibitive when the sample size $N$ grows. Especially, updating location parameters via \eqref{eq:EM_location} is problematic since the membership matrix $\bfGamma$ is highly likely to be dense and a weighted \Frechet median over a large number of observations needs to be computed. We present two popular heuristics to reduce the sample size - hard and stochastic assignments - that directly manipulate $\bfGamma$ so that the pertained computation is limited to a smaller subset of the sample. Denote the $n$-th row of $\bfGamma$ as $\bfGamma_n$ and $\textsf{sample}(\mathbf{v}_z, \textrm{probability}=\mathbf{p}_z)$ is a sampling procedure to randomly draw an element from a vector  $\mathbf{v}_z$ with probability $\mathbf{p}_z$.  Two  assignments are done as follows,
\begin{eqnarray}
\textsf{hard}(\gamma_{nk}) &=& \begin{cases}
1, &\textrm{if $k$ is maximal index of $\bfGamma_n$,}\\0, &\textrm{otherwise}.
\end{cases} \label{eq:heuristic_hard}	\\
\textsf{stochastic}(\gamma_{nk}) &=& \begin{cases}
1, &\textrm{if $k=\textsf{sample}(1:K, ~\textrm{probability}=\bfGamma_n)$},\\
0, &\textrm{otherwise}. \label{eq:heuristic_stochastic}
\end{cases}
\end{eqnarray}
The hard assignment is equivalent to predict discrete-valued label for an observation, while the stochastic assignment performs the same task in a probabilistic manner. Given a large sample, these sparsification strategies help to reduce execution time and space complexity in updating both location and scale parameters. We note that the hard assignment is optimal in the sense that a lower bound of the incomplete-data log-likelihood is maximized \cite{banerjee_2005_ClusteringUnitHypersphere}.

Third, we describe some practical details from an implementation side. We initialize the algorithm using partition structure obtained from $k$-means clustering \cite{macqueen_1967_MethodsClassificationAnalysis}. This is because most computing platforms are equipped with efficient, robust implementations of the algorithm. Besides, a large amount of geometric information is preserved by an equivariant embedding \cite{bhattacharya_2012_NonparametricInferenceManifolds}, which is the identity map on $\bbS^p$, in that this approach preserves certain degree of geometric information. Another practical issue is to determine convergence of the algorithm. It is natural to terminate iterations if log-likelihood stops to increase. From a computational perspective, however, the log-likelihood is not an appealing option since computing the quantity requires density evaluation with updated parameters at each iteration while no intermediate quantities may be used. Hence, we suggest to use incremental change in a soft clustering matrix $\|\bfGamma^{(t+1)} - \bfGamma^{(t)}\|$ as a quantity to terminate iterations since it implies that clustering results do not evolve and no extra updates are needed.

We close this section by considering a potential variant of our model. The finite mixture model has long been known to be closely related to the $k$-means algorithm, which has been also observed in the context of directional statistics  \cite{dhillon_2001_ConceptDecompositionsLarge, banerjee_2005_ClusteringUnitHypersphere, bishop_2006_PatternRecognitionMachine}. Assuming the homogeneous model of equal scale, an entry of membership matrix $\bfGamma$ is written as 
\begin{eqnarray}
\gamma_{nk} = \frac{\pi_k^{(t)} \fSL(\bfx_n~|~\bfmu_k^{(t)}, \sigma_k^{(t)})}{\sum_{j=1}^K \pi_j^{(t)} \fSL(\bfx_n~|~\bfmu_j^{(t)}, \sigma_j^{(t)})} = 
\frac{\pi_k \exp \left( -{d(\bfx_n, \bfmu_k)}/{\sigma} \right) }{\sum_{j=1}^K \pi_j \exp \left( -{d(\bfx_n, \bfmu_j)}/{\sigma} \right)},
\end{eqnarray}
which quantifies likelihood for the $n$-th observation to belong to the $k$-th component or cluster. Let $k^*$ be an index for $d(\bfx_n, \bfmu_k)$ being the smallest. When $\sigma \rightarrow 0$, the term $d(\bfx_n, \bfmu_{k^*})$ decays at the slowest rate in the denominator than others so that all other $\gamma_{nk}$'s approach to zero. This procedure is identical to how the $k$-medians clustering algorithm updates assignment of each observation. We can validate this reasoning by taking a limit of the complete-data log-likelihood
\begin{eqnarray}
\rmE_{\bfZ|\bfX, \bfTheta} \lbrack \log \rmP(\bfX,\bfZ|\bfTheta) \rbrack  \xrightarrow{\sigma \rightarrow 0} - \sum_{n=1}^N \sum_{k=1}^K \gamma_{nk} d(\bfx_n, \bfmu_k) + \textrm{some constant}.
\end{eqnarray}
Therefore, in the limit sense, maximizing the expected complete-data log-likelihood amounts to minimization of cost function in a generic $k$-medians problem, which has not been popularized in directional statistics up to our knowledge.

\begin{algorithm}[ht]
	\caption{EM algorithm for the mixture of SL distributions.}
	\label{code:em_splaplace}
	\begin{algorithmic}
		\REQUIRE a random sample $\bfx_1, \ldots, \bfx_N \in \bbS^p$, number of clusters $K$.
		\ENSURE a soft clustering/membership matrix $\bfGamma$.		
		\STATE Initialize $ \bfTheta^{(0)} = \lbrace \pi_k^{(0)}, \bfmu_k^{(0)}, \sigma_k^{(0)}\rbrace_{k=1}^K$.
		\REPEAT 
		\STATE \{E-step\}
		\FOR{$n=1:N$}
		\FOR {$k=1:K$ }
		\STATE $\bfGamma (n,k)= \pi_k^{(t)} \fSL (\bfx_n~\vert~ \bfmu_k^{(t)}, \sigma_k^{(t)})$
		\ENDFOR
		\STATE $\bfGamma(n,:) = \bfGamma(n,:)/ \sum_{k=1}^K \bfGamma(n,k)$
		\ENDFOR 
		\STATE \{Heuristics\}
		\IF {hard assignment}
		\STATE $\bfGamma \leftarrow \textsf{hard}(\bfGamma)$ by \eqref{eq:heuristic_hard}.
		\ELSIF {stochastic assignment}
		\STATE $\bfGamma \leftarrow \textsf{stochastic}(\bfGamma)$ by \eqref{eq:heuristic_stochastic}.
		\ENDIF
		\STATE \{M-step\}
		\FOR{$k=1:K$}
		\STATE $\pi_k^{(t+1)} = \sum_{n=1}^N \gamma_{nk} / N$.
		\STATE $\bfmu_k^{(t+1)} = \underset{\bfmu \in \bbS^p}{\argmin~}  \sum_{n=1}^N \gamma_{nk} \cdot d (\bfx_n, \bfmu)$ by Algorithm \ref{code:algorithm_location}. 
		\ENDFOR
		\IF {homogeneous model}
		\STATE Update $\sigma^{(t+1)}$ using \eqref{eq:EM_scale_homo} by Algorithm \ref{code:algorithm_scale}.
		\ELSE
		\FOR {$k=1:K$}
		\STATE  Update $\sigma_k^{(t+1)}$ using \eqref{eq:EM_scale_hetero} by Algorithm \ref{code:algorithm_scale}. 
		\ENDFOR
		\ENDIF
		\UNTIL convergence.
	\end{algorithmic}
\end{algorithm}

\section{Experiments}\label{sec6:experiments}

We come to assess how the proposed algorithms perform for the tasks of parameter estimation and model-based clustering. The first experiment is composed of evaluating parameter estimation algorithms with simulated data. Then, we investigate effectiveness of model-based clustering with the finite mixture of SL distributions using simulated and real data examples.

\subsection{Estimating parameters}

We evaluate performance of the proposed algorithms for maximum likelihood estimation of two parameters. We start by presenting  the simplest case where two fixed parameters $\bfmu_0 \in (1,0,0,0,0,0)^\top \in \bbS^{5}$ and $\sigma_0 = 0.1$ are used to generate random samples of varying size from 25 to $475$. Each setting is run 100 times and empirical distributions of two performance measures, run time and accuracy, are reported. For numerical optimization, the stopping criterion of $\epsilon = 10^{-8}$ was used throughout all computations, which is approximately equal to a square root of the machine epsilon in double precision. 

\begin{figure}[h]
	\centering
	\includegraphics[width=0.95\linewidth]{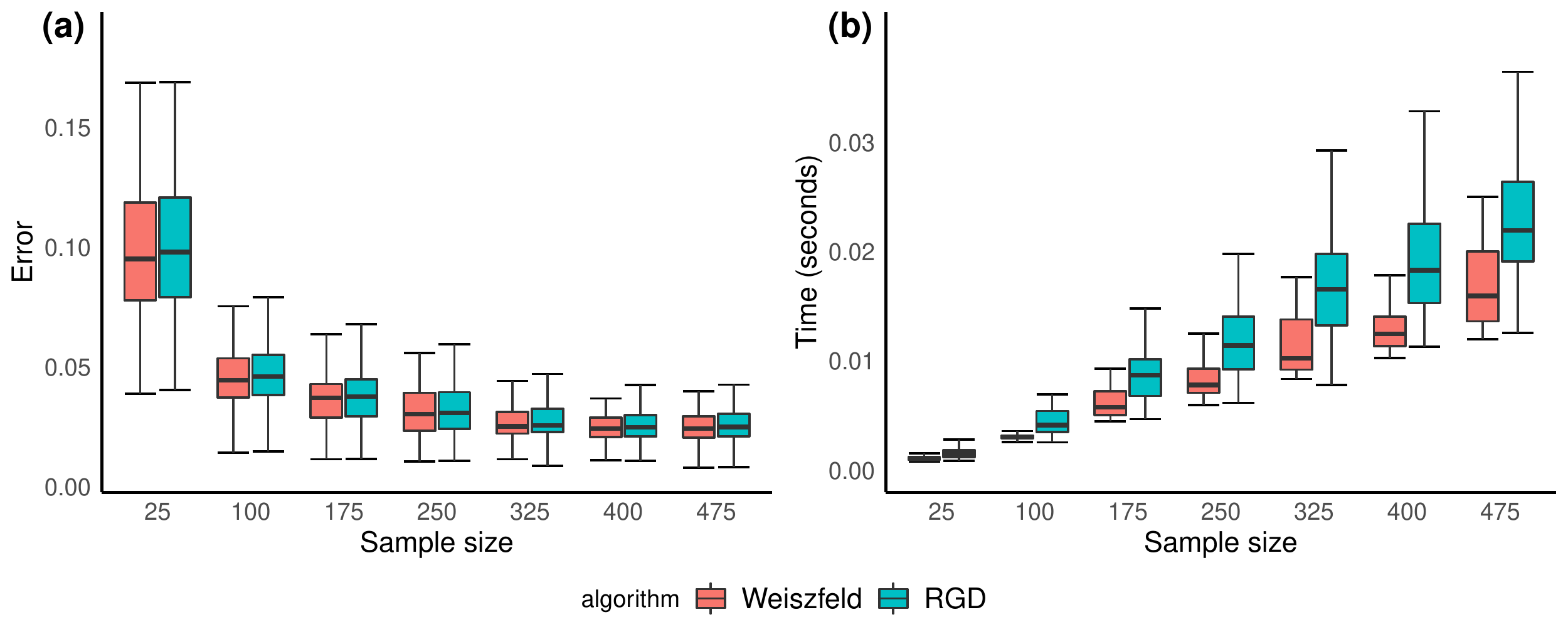}
	\caption{Empirical distribution of performance measures, (a) accuracy and (b) elapsed time, for the location estimation problem. At each run, a random sample from the SL distribution of parameters $\bfmu_0 = (1,0,0,0,0,0)^\top, \sigma_0 = 0.1$ was drawn and the procedure was repeated 100 times. }
	\label{fig:estimation_location}
\end{figure}

For the location estimation problem, we compared the geometric Weiszfeld method as shown in Algorithm \ref{code:algorithm_location} and the standard version of Riemannian gradient descent (RGD) algorithm. The latter relies on line search to determine how much an iterate shall proceed along the descent direction. Given a random sample $\bfX = \{\bfx_1, \ldots, \bfx_n\} \subset \bbS^5$ and a maximum likelihood estimate $\hat{\bfmu}_{\mathrm{MLE}}$, accuracy is measured by the geodesic distance $d(\hat{\bfmu}_{\textrm{MLE}}, \bfmu_0) = \cos^{-1}(\langle \hat{\bfmu}_{\textrm{MLE}}, \bfmu_0\rangle)$. Two performance measures under the setting are shown in \figurename \ref{fig:estimation_location}. As the sample size grows, it is seen that better estimates are obtained in both methods at almost parallel levels of accuracy. In terms of computational cost, however, the Weiszfeld algorithm consistently outperforms the RGD and the degree of disparity gets larger along the growing sample size. This is an expected phenomenon in the sense that line search requires repeated evaluation of the cost functional while the Weiszfeld algorithm requires a single evaluation at each iteration. Therefore, the RGD may converge in fewer iterations than the Weiszfeld algorithm at the increased cost of overall computation. 

\begin{figure}[ht]
	\centering
	\includegraphics[width=0.95\linewidth]{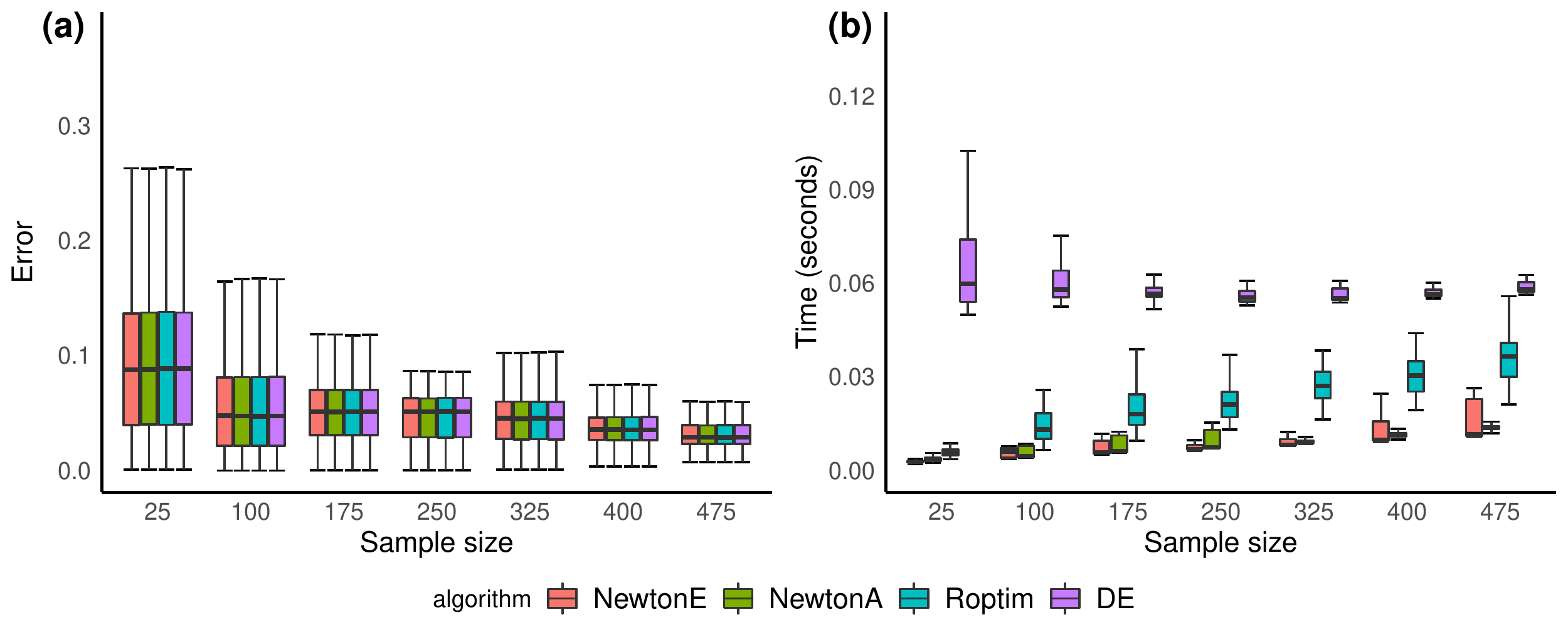}
	\caption{Empirical distribution of performance measures, (a) accuracy and (b) elapsed time, for the scale estimation problem. Each run drew a random sample from the SL distribution parametrized by $\bfmu_0 = (1,0,0,0,0,0)^\top, \sigma_0 = 0.1$. The procedure was repeated 100 times. }
	\label{fig:estimation_scale}
\end{figure}

Similarly, we compared performance of the proposed Newton's methods against two univariate optimization routines that do not require derivative information. Our first choice is a default optimization routine in \textsf{R} programming language \cite{rcoreteam_2022_LanguageEnvironmentStatistical} that uses golden-section search with successive parabolic interpolation, which we will denote as \textsf{Roptim}. Another is the differential evolution algorithm for global optimization using a heuristic approach \cite{storn_1997_DifferentialEvolutionSimple} and will be denoted as \textsf{DE}. We derived two updating rules for the Newton's methods using exact integral evaluation and approximation of the derivatives using finite difference, which will be called as \textsf{NewtonE} and \textsf{NewtonA}, respectively. We measured accuracy of the estimate  $\hat{\sigma}_{\textrm{MLE}}$ by relative error $|\hat{\sigma}_{\textrm{MLE}} - \sigma_0|/\sigma_0$. Performance measures for all 4 algorithms under the aforementioned setting are summarized in \figurename \ref{fig:estimation_scale}. The quality of estimation was comparable across all methods, which is not surprising since all algorithms start from an identical location estimate. When it comes to computational efficiency, however, both our proposed algorithms show superior performance to the others.

Besides the simple case, we performed  extensive experiments to compare performance of the proposed algorithms across different settings. From the tabularized results that are reported in the Supplementary Materials, we verified the consistent patterns across heterogeneous experimental conditions regarding the accuracy and elapsed time.

\subsection{Clustering}

We consider two examples for model-based clustering of data on the unit hypersphere in order to demonstrate effectiveness of the SL mixture model. In both examples, we compare mixtures of SL distributions with soft (\textsc{moSL-soft}) and hard (\textsc{moSL-hard}) assignment with a number of competing algorithms including $k$-means (\textsc{kmeans}) \cite{macqueen_1967_MethodsClassificationAnalysis}, spherical $k$-means (\textsc{spkmeans}) \cite{dhillon_2001_ConceptDecompositionsLarge}, mixture of vMF distributions (\textsc{movMF}) \cite{banerjee_2005_ClusteringUnitHypersphere}, and mixture of SN distributions (\textsc{moSN}) \cite{you_2022_ParameterEstimationModelbased}. In order to evaluate quality of algorithms, clustering comparison indices between a true or desired label and an attained label are used. Among many alternatives, we report Jaccard index \cite{jaccard_1912_DISTRIBUTIONFLORAALPINE}, Rand index \cite{rand_1971_ObjectiveCriteriaEvaluation}, and normalized mutual information (NMI) \cite{strehl_2002_ClusterEnsemblesKnowledge} since these have values in $[0,1]$. Furthermore, their numeric values represent the degree of similarity between two clusterings in a way that the higher an index value is, the more agreement is present. Another note is that three indices are invariant to relabeling of assignments.

First, we use simulated data according to the \textsf{small-mix} example from \cite{banerjee_2005_ClusteringUnitHypersphere} with minor modification as adopted in \cite{you_2022_ParameterEstimationModelbased}. The data generating model in this example is a mixture of two SN distributions on $\bbS^1 \subset \bbR^2$. Two equally weighted components are parametrized by $(\bfmu_1, \lambda_1) = ([-0.251,-0.968], 10)$ and $(\bfmu_2, \lambda_2) = ([0.399, 0.917], 2)$ with heterogeneous dispersions. In each run, a total of 200 observations are randomly drawn from the model.

\begin{figure}[h]
	\centering
	\includegraphics[width=0.95\linewidth]{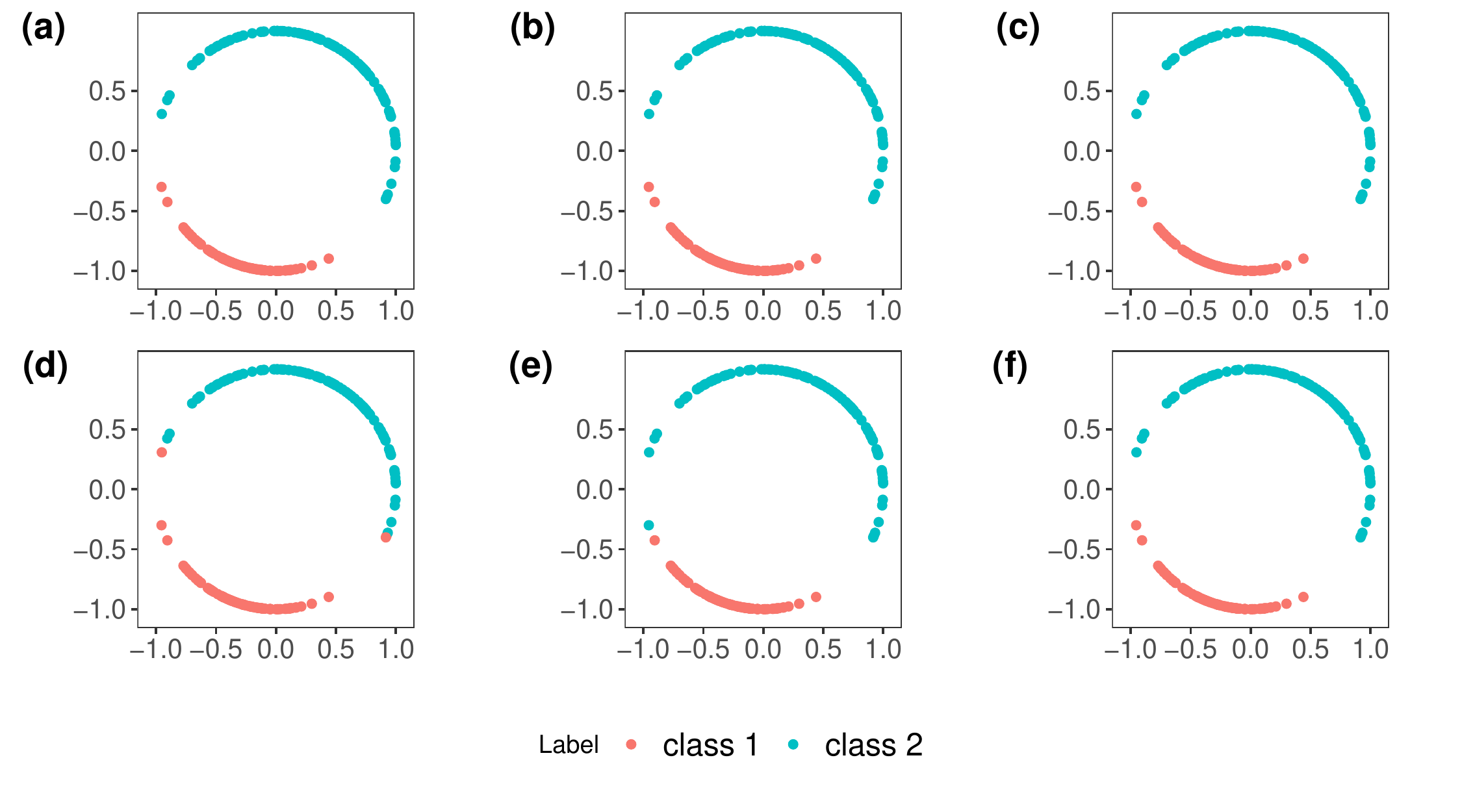}
	\vspace{-.2cm}
	\caption{Visualization of the \textsf{small-mix} example data consisting of two distinct classes on $\bbS^1$. The true label is shown in (a) and clustering results are given from the algorithms as follows; (b) \textsc{moSL}, (c) \textsc{kmeans}, (d) \textsc{spkmeans}, (e) \textsc{movMF}, and (f) \textsc{moSN} when the number of clusters is set as 2.}
	\label{fig:simulation_smallmix}
\end{figure}

We show a random sample with true label and clustering results in Fig. \ref{fig:simulation_smallmix}, where the true label is perfectly recovered from our proposed mixture of SL distributions along with $k$-means and SN mixture models. It is not surprising that the standard $k$-means algorithm achieves the perfect clustering since the sample has two distinct components. It was also reported in \cite{you_2022_ParameterEstimationModelbased} that the SN mixture outperforms spherical $k$-means and vMF mixture models, both of which misclassified observations near the boundary. We repeated the test 100 times and compared average clustering indices, which are summarized in Table \ref{table:small_mix}. When the true number of clusters is set as $K=2$, we witnessed an interesting phenomenon that the SL mixture is the best performing method even though the data was generated from the mixture of SN components. One  explanation of this phenomenon is related to robust estimation of cluster means for the SL distribution so that the perturbed observations at the boundary of two components less affect the overall identification of clusters. When the models are misspecified, the SL mixture returned smaller clustering comparison indices than other mixture models. Since the true partition is not known in most real cases, we consider this as an encouraging pattern since a valid mixture model should discourage misspecification.

\begin{table}[h]
	\caption{Average of clustering quality indices from 100 runs for the \textsf{small-mix} experiment. For each column, the cell containing bold-face numeric indicates that the corresponding row is the best performing model given the cluster number $K$ and quality index.}
	\label{table:small_mix}
	\vskip-0.3cm
	\smallskip
	\centering\small
	\begin{tabular}{cccccccccc}
		\hline
		\multirow{2}{*}{} & \multicolumn{3}{c}{Jaccard} & \multicolumn{3}{c}{Rand} & \multicolumn{3}{c}{NMI} \\ \cline{2-10} 
		& $K=2$     & $K=3$    & $K=4$    & $K=2$      & $K=3$     & $K=4$     & $K=2$   & $K=3$   & $K=4$    \\ \hline
		\textsc{moSL-soft} & \textbf{0.9862} & 0.7521 & 0.6107 & \textbf{0.9930} & 0.8756 & 0.8052 & \textbf{0.9712} & 0.7972 & 0.7194 \\ 
		\textsc{moSL-hard} & 0.9689 & 0.7430 & 0.5861 & 0.9841 & 0.8708 & 0.7926 & 0.9422 & 0.7848 & 0.7015 \\ 
		\textsc{kmeans} & 0.9575 & 0.7463 & 0.5588 & 0.9782 & 0.8726 & 0.7789 & 0.9270 & 0.7822 & 0.6930 \\ 
		\textsc{spkmeans} & 0.9465 & 0.7427 & 0.5997 & 0.9724 & 0.8709 & 0.7996 & 0.9118 & 0.7935 & 0.7104 \\ 
		\textsc{movMF} & 0.9852 & \textbf{0.7927} & \textbf{0.6841} & 0.9825 & \textbf{0.8963} & \textbf{0.8424} & 0.9607 & 0.8193 & \textbf{0.7582} \\ 
		\textsc{moSN} & 0.9853 & 0.7907 & 0.6325 & 0.9830 & 0.8953 & 0.8167 & 0.9627 & \textbf{0.8225} & 0.7335 \\  \hline
	\end{tabular}
\end{table}

We now turn to the real data example using \textsf{household} data, which is a survey data of household-level expenditures on several commodity groups among 40 individuals (20 males, 20 females) \cite{everitt_2010_HandbookStatisticalAnalyses}. Following the convention of \cite{hornik_2014_MovMFPackageFitting}, we extracted expenditures of three categories - food, housing, and service. Our interest is on the proportion of each category so that a profile vector is obtained by $\ell_1$ normalization and projected onto $\bbS^2$ by square-root transformation, which is a common practice in studying compositional data \cite{marron_2021_ObjectOrientedData}.

\begin{table}[]
	\caption{Clustering quality indices where the number of clusters varies for the \textsf{household} data projected onto $\bbS^2$ by square-root transformation of compositional data. For each column, the cells with bold-face numerics indicate that the corresponding row is 
		the best performing algorithm given the cluster number $K$ and quality index.}
	\label{table:household}
	\vskip-0.3cm
	\smallskip
	\centering\small
	\begin{tabular}{cccccccccc}
		\hline
		\multirow{2}{*}{} & \multicolumn{3}{c}{Jaccard} & \multicolumn{3}{c}{Rand} & \multicolumn{3}{c}{NMI} \\ \cline{2-10} 
		& $K=2$     & $K=3$    & $K=4$    & $K=2$      & $K=3$     & $K=4$     & $K=2$   & $K=3$   & $K=4$    \\ \hline
		\textsc{moSL-soft} & \textbf{1.0000} & 0.7275 & 0.6342 & \textbf{1.0000} & 0.8603 & \textbf{0.8218} & \textbf{1.0000} & 0.7244 & \textbf{0.7524} \\ 
		\textsc{moSL-hard} & 0.5920 & 0.7275 & 0.5363 & 0.7385 & 0.8603 & 0.7705 & 0.5105 & 0.7244 & 0.6546 \\ 
		\textsc{kmeans} & 0.5920 & \textbf{0.7789} & 0.5363 & 0.7385 & \textbf{0.8923} & 0.7705 & 0.5105 & \textbf{0.8331} & 0.6546 \\ 
		\textsc{spkmeans} & 0.5920 & \textbf{0.7789} & 0.5363 & 0.7385 & \textbf{0.8923} & 0.7705 & 0.5105 & \textbf{0.8331} & 0.6546 \\ 
		\textsc{movMF} & 0.9025 & 0.7275 & \textbf{0.6450} & 0.9500 & 0.8603 & 0.8179 & 0.8558 & 0.7244 & 0.6645 \\ 
		\textsc{moSN} & 0.5920 & 0.7275 & 0.5363 & 0.7385 & 0.8603 & 0.7705 & 0.5105 & 0.7244 & 0.6546 \\  \hline
	\end{tabular}
\end{table}

\begin{figure}[h]
	\centering
	\includegraphics[width=0.95\linewidth]{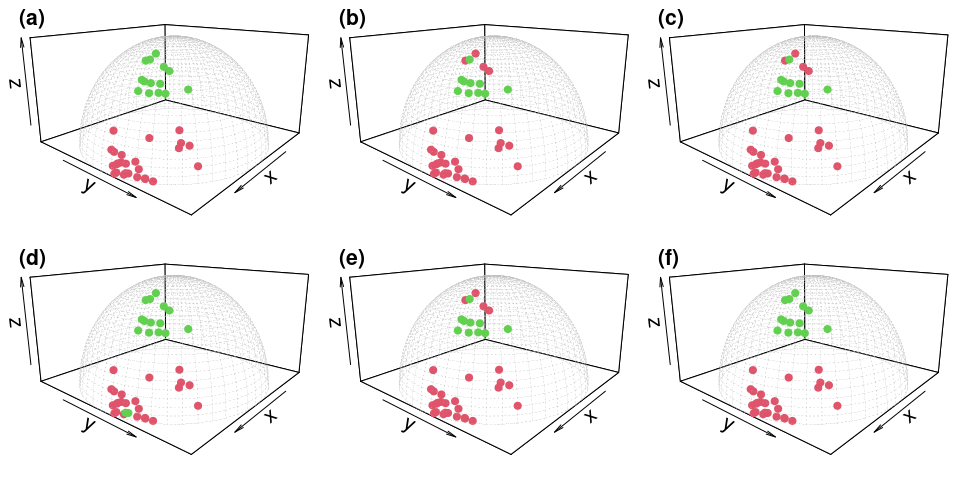}
	\vspace{-.2cm}
	\caption{
		Visualization of the \textsf{household} data on $\bbS^2$. The true label defined by gender is shown in (a) and clustering results are given from the algorithms as follows; (b) \textsc{kmeans}, (c) \textsc{spkmeans}, (d) \textsc{movMF}, (e) \textsc{moSN}, and (f) \textsc{moSL}. All models for comparison are specified with the number of clusters as 2.}
	\label{fig:clust_household}
\end{figure}

Similar to the \textsf{small-mix} example, we compared performance of 6 different algorithms for varying number of clusters with the projected \textsf{household} data. Quality of clustering performance was measured by regarding the gender label as ground-truth cluster label in that the true number of clusters is 2. Results are summarized in Table \ref{table:household}.  While the SL and vMF mixtures turned out to be the only models that recovered gender-separated clusters, it is worth to mention that our proposed SL mixture model with standard soft membership showed perfect clustering results when model was correctly specified. Comparative visualization of different clustering outputs is presented in Fig. \ref{fig:clust_household}. Contrary to the SL mixture that perfectly recovers the gender-separated classes, other methods show somewhat strange patterns. For example, both $k$-means and spherical $k$-means misclassified observations near the north pole as belonging to the same class with those near the equator. This may come from the fact that these algorithms do not acknowledge much for the  constrained geometry of the domain. This pattern is also witnessed in two mixture models of vMF and SN components that the distinct classes are seemingly contaminated by farthest points. This may be caused by the fact that data from the male group is much more dispersed than that of the female group. When the SL distribution was fitted separately, the females' expenditure data was distributed with $\hat{\sigma}_{\textrm{MLE}} = 0.0643$, which is much smaller than that of the male group with $\hat{\sigma}_{\textrm{MLE}} = 0.1426$. Since the scale parameter $\sigma$ amounts to variability within the data, this means that the male group, compared to the female group, is less concentrated and shows higher degree of heterogeneity. In reality, it is nearly impenetrable to distinguish presence of outliers from noise of substantial magnitude. Therefore, specifying a mixture model with robust components can help to deal with the scenario that we observed. 

\section{Conclusions}\label{sec7:conclusion}

We proposed the SL distribution, a generalization of the Laplace distribution onto the unit hypersphere. The distribution was shown to be well-defined and provided with a sampling scheme, which is a fundamental tool arising in many computational pipelines. We showed that the maximum likelihood estimates of the governing parameters uniquely exist under mild conditions on the support of an empirical measure.  We also proposed  numerical optimization routines for maximum likelihood estimation of location and scale parameters since the SL distribution does not admit closed-form formulae for the estimates. The proposed algorithm is validated with extensive experiments using simulated data. An application of our proposal is model-based clustering of spherical data under the finite mixture framework where the components are SL distributions. Our experiments with simulated and real data showed that there is gain from the SL mixture  where a large amount of noise is suspected. 

We close by discussing several topics of interests for future studies. As noted before, the current sampling scheme built upon rejection sampler could be inefficient when a scale parameter is set to be very small. We expect that further investigation along this direction makes the proposed distribution a more appealing component not only for statistical inference but also for probabilistic deep learning such as variational autoencoders. Another line of research is reserved for refining clustering procedures. While our current exploration focuses on the finite mixture framework, availability of an effective sampler helps to open up opportunities such as a nonparametric Bayesian density estimation \cite{bhattacharya_2012_NonparametricBayesClassification} and modelling hierarchical and temporal data on the unit hypersphere  \cite{gopal_2014_MisesfisherClusteringModels}. The last direction is anisotropic extension of the SL distribution as done in \cite{hauberg_2018_DirectionalStatisticsSpherical} for the SN distribution, which will be beneficial in flexible modelling of directionally dependent data on the unit hypersphere.

\section*{Appendix}\label{sec:appendix}
We describe technical details on derivation of rejection sampler for the SL distribution. A standard rejection sampler aims at drawing random samples from a target density $f(\bfx)$ by samples from a proposal distribution $g(\bfx)$, assuming the likelihood ratio $f(\bfx)/g(\bfx)$ is upper bounded by some constant $M \in (1, \infty)$ over the entire domain. A draw from $g(\bfx)$ is accepted as a random sample from $f(\bfx)$ with probability $f(\bfx)/M g(\bfx)$ and the process is repeated until a successful draw. For the SL distribution with parameters $(\bfmu,\sigma) \in \bbS^p \times \bbRpos$, we consider the SN distribution of parameters $(\bfmu, 1/\sigma) \in \bbS^p \times \bbRpos$ as a proposal density. We note that density functions for SN and SL distributions are explicitly written as
\begin{equation*}
\fSN(\bfx~|~\bfmu,\lambda) = \frac{1}{Z_p(\lambda)} \exp \left(-\frac{\lambda}{2} d^2(\bfx,\bfmu)\right)\quad\text{and}\quad \fSL(\bfx~|~\bfmu,\sigma) = \frac{1}{C_p(\sigma)} \exp \left(-\frac{1}{\sigma} d(\bfx,\bfmu)\right),
\end{equation*}
for some normalizing constant $Z_p$ for the SN distribution \cite{hauberg_2018_DirectionalStatisticsSpherical}. Under the scenario, the likelihood ratio is given by
\begin{align*}
\frac{\fSL(\bfx~|~\bfmu,\sigma)}{\fSN(\bfx~|~\bfmu,\lambda)} &= 
\frac{Z_p (\lambda)}{C_p (\sigma)} \exp\left(-\frac{1}{\sigma}d(\bfx,\bfmu) + \frac{\lambda}{2} d^2(\bfx,\bfmu)\right)\\ 
&=
\frac{Z_p (1/\sigma)}{C_p (\sigma)} \exp\left(-\frac{1}{\sigma}d(\bfx,\bfmu) + \frac{1}{2\sigma} d^2(\bfx,\bfmu)\right)\\
&= \frac{Z_p (1/\sigma)}{C_p (\sigma)} \exp\left(\frac{(d(\bfx,\bfmu)-1)^2}{2\sigma}\right) \cdot  \exp\left(-\frac{1}{2\sigma}\right)\\  &\leq \frac{Z_p (1/\sigma)}{C_p (\sigma)} \exp\left(\frac{((\pi-1)^2}{2\sigma}\right) \cdot \exp\left(-\frac{1}{2\sigma}\right) =: M,
\end{align*}
where the constant $M$ is an upper bound of the ratio. An inequality comes from the fact that injectivity radius of a unit hypersphere is $\pi$, i.e., any two points on the unit hypersphere have distance of $\pi$ at most. Then, the corresponding probabilistic acceptance threshold $\tau$ is as follows;
\begin{align*}
\tau &= \frac{\fSL(\bfx~|~\bfmu,\sigma)}{M \fSN(\bfx~|~\bfmu,1/\sigma)}\\
&= \frac{C_p(\sigma)}{Z_p (1/\sigma)} 
\exp\left(-\frac{(\pi-1)^2}{2\sigma}\right) \cdot  \exp\left(\frac{1}{2\sigma}\right)\cdot 
\frac{1}{C_p(\sigma)} \exp\left(-\frac{d(\bfx,\bfmu)}{\sigma}\right) \cdot  Z_p (1/\sigma) \cdot  \exp\left(\frac{1}{2\sigma} d^2 (\bfx,\bfmu) \right)\\
&= \exp\left(\frac{1}{2\sigma}d^2 (\bfx,\bfmu) - \frac{1}{\sigma}d (\bfx,\bfmu) + \frac{1}{2\sigma}\right)\cdot \exp\left(-\frac{(\pi-1)^2}{2\sigma}\right)\\ &= \exp\left(\frac{(d(\bfx,\bfmu)-1)^2 - (\pi-1)^2}{2\sigma} \right).
\end{align*}

\bibliographystyle{dcu}
\bibliography{splaplace}

\newpage
\section*{Supplementary Material}\label{sec:supp}
\begin{center}
	\begin{longtable}{LLLLLLL}
		\caption{Average accuracy and elapsed time for the location parameter estimation example from 100 repeats per setting. Accuracy is measured by the geodesic distance $d(\hat{\mathbb{\mu}}_{\text{MLE}}, \mathbf{\mu}_0)$ and elapsed time is measured in seconds. }  
		\label{table:extended_location}\\
		\hline
		\multirow{2}{*}{dimension}          & \multirow{2}{*}{$\sigma_0$} & \multirow{2}{*}{$n$} & \multicolumn{2}{c}{ Weiszfeld} & \multicolumn{2}{c}{RGD} \\ \cline{4-7} 
		&                     &    & accuracy & time & accuracy & time \\ \hhline{-------}
		\multirow[c]{28}{*}{$p=5$} 
		& \multirow{4}{*}{0.01}&50 & 0.00672 & 0.00031 & 0.00693 & 0.00046 \\ \cline{3-7} 
		&&100 & 0.00445 & 0.00053 & 0.00457 & 0.00075 \\ \cline{3-7} 
		&&250 & 0.00277 & 0.00125 & 0.00284 & 0.00177 \\ \cline{3-7} 
		&&500 & 0.00204 & 0.00227 & 0.00209 & 0.00340 \\ \cline{2-7} 
		& \multirow{4}{*}{0.05}&50 & 0.03223 & 0.00029 & 0.03301 & 0.00042 \\ \cline{3-7} 
		&&100 & 0.02189 & 0.00059 & 0.02258 & 0.00090 \\ \cline{3-7} 
		&&250 & 0.01318 & 0.00141 & 0.01349 & 0.00206 \\ \cline{3-7} 
		&&500 & 0.00925 & 0.00243 & 0.00949 & 0.00356 \\ \cline{2-7} 
		& \multirow{4}{*}{0.1}&50 & 0.06844 & 0.00032 & 0.07044 & 0.00044 \\ \cline{3-7} 
		&&100 & 0.04683 & 0.00062 & 0.04803 & 0.00086 \\ \cline{3-7} 
		&&250 & 0.03116 & 0.00140 & 0.03195 & 0.00196 \\ \cline{3-7} 
		&&500 & 0.02404 & 0.00264 & 0.02471 & 0.00400 \\ \cline{2-7} 
		& \multirow{4}{*}{0.5}&50 & 0.33819 & 0.00086 & 0.34528 & 0.00116 \\ \cline{3-7} 
		&&100 & 0.22558 & 0.00114 & 0.23096 & 0.00167 \\ \cline{3-7} 
		&&250 & 0.13930 & 0.00276 & 0.14277 & 0.00401 \\ \cline{3-7} 
		&&500 & 0.08994 & 0.00519 & 0.09231 & 0.00777 \\ \cline{2-7} 
		& \multirow{4}{*}{1}&50 & 0.56018 & 0.00102 & 0.57682 & 0.00139 \\ \cline{3-7} 
		&&100 & 0.44617 & 0.00199 & 0.45898 & 0.00285 \\ \cline{3-7} 
		&&250 & 0.27722 & 0.00441 & 0.28263 & 0.00662 \\ \cline{3-7} 
		&&500 & 0.19769 & 0.00862 & 0.20139 & 0.01300 \\ \cline{2-7} 
		& \multirow{4}{*}{5}&50 & 1.22967 & 0.00152 & 1.26201 & 0.00220 \\ \cline{3-7} 
		&&100 & 1.09516 & 0.00338 & 1.12680 & 0.00483 \\ \cline{3-7} 
		&&250 & 0.82620 & 0.00891 & 0.84687 & 0.01346 \\ \cline{3-7} 
		&&500 & 0.73339 & 0.01674 & 0.75288 & 0.02417 \\ \cline{2-7} 
		& \multirow{4}{*}{10}&50 & 1.13413 & 0.00145 & 1.16390 & 0.00207 \\ \cline{3-7} 
		&&100 & 1.15891 & 0.00366 & 1.19725 & 0.00559 \\ \cline{3-7} 
		&&250 & 0.97991 & 0.00864 & 1.00427 & 0.01178 \\ \cline{3-7} 
		&&500 & 0.87189 & 0.01705 & 0.88945 & 0.02503 \\ \hline
		\multirow[c]{28}{*}{$p=10$}
		& \multirow{4}{*}{0.01}&50 & 0.01113 & 0.00026 & 0.01139 & 0.00040 \\ \cline{3-7} 
		&&100 & 0.00796 & 0.00043 & 0.00818 & 0.00060 \\ \cline{3-7} 
		&&250 & 0.00502 & 0.00113 & 0.00515 & 0.00171 \\ \cline{3-7} 
		&&500 & 0.00348 & 0.00216 & 0.00357 & 0.00311 \\ \cline{2-7} 
		& \multirow{4}{*}{0.05}&50 & 0.05798 & 0.00029 & 0.05930 & 0.00041 \\ \cline{3-7}
		&&100 & 0.04115 & 0.00051 & 0.04210 & 0.00069 \\ \cline{3-7}
		&&250 & 0.02366 & 0.00140 & 0.02434 & 0.00198 \\ \cline{3-7}
		&&500 & 0.01828 & 0.00244 & 0.01878 & 0.00363 \\ \cline{2-7}
		& \multirow{4}{*}{0.1}&50 & 0.12417 & 0.00036 & 0.12668 & 0.00051 \\ \cline{3-7}
		&&100 & 0.08499 & 0.00077 & 0.08700 & 0.00116 \\ \cline{3-7}
		&&250 & 0.05368 & 0.00159 & 0.05505 & 0.00246 \\ \cline{3-7}
		&&500 & 0.03857 & 0.00336 & 0.03967 & 0.00499 \\ \cline{2-7}
		& \multirow{4}{*}{0.5}&50 & 0.64652 & 0.00099 & 0.66424 & 0.00139 \\ \cline{3-7}
		&&100 & 0.43624 & 0.00214 & 0.44816 & 0.00305 \\ \cline{3-7}
		&&250 & 0.29295 & 0.00635 & 0.29953 & 0.00917 \\ \cline{3-7}
		&&500 & 0.20830 & 0.00949 & 0.21271 & 0.01310 \\ \cline{2-7}
		& \multirow{4}{*}{1}&50 & 0.92133 & 0.00152 & 0.94644 & 0.00224 \\ \cline{3-7}
		&&100 & 0.72675 & 0.00308 & 0.74565 & 0.00444 \\ \cline{3-7}
		&&250 & 0.50497 & 0.00815 & 0.51890 & 0.01147 \\ \cline{3-7}
		&&500 & 0.36499 & 0.01601 & 0.37349 & 0.02377 \\ \cline{2-7}
		& \multirow{4}{*}{5}&50 & 1.34914 & 0.00164 & 1.39048 & 0.00243 \\ \cline{3-7}
		&&100 & 1.32430 & 0.00396 & 1.35234 & 0.00588 \\ \cline{3-7}
		&&250 & 1.27650 & 0.00999 & 1.31860 & 0.01441 \\ \cline{3-7}
		&&500 & 1.23680 & 0.02009 & 1.27001 & 0.02874 \\ \cline{2-7}
		& \multirow{4}{*}{10}&50 & 1.36467 & 0.00159 & 1.39984 & 0.00229 \\ \cline{3-7}
		&&100 & 1.37816 & 0.00389 & 1.40600 & 0.00547 \\ \cline{3-7}
		&&250 & 1.32753 & 0.01000 & 1.35787 & 0.01466 \\ \cline{3-7}
		&&500 & 1.23180 & 0.01968 & 1.25773 & 0.02876 \\ \hline
		\multirow[c]{28}{*}{$p=20$}
		& \multirow{4}{*}{0.01}& 50 & 0.01796 & 0.00056 & 0.01828 & 0.00088 \\ \cline{3-7}
		&&   100 & 0.01295 & 0.00090 & 0.01328 & 0.00132 \\ \cline{3-7}
		&&   250 & 0.00780 & 0.00201 & 0.00797 & 0.00288 \\ \cline{3-7}
		&&   500 & 0.00555 & 0.00410 & 0.00567 & 0.00608 \\ \cline{2-7}
		& \multirow{4}{*}{0.05}&   50 & 0.09916 & 0.00091 & 0.10187 & 0.00132 \\ \cline{3-7}
		&&   100 & 0.06678 & 0.00143 & 0.06853 & 0.00200 \\ \cline{3-7}
		&&   250 & 0.04320 & 0.00291 & 0.04438 & 0.00410 \\ \cline{3-7}
		&&   500 & 0.02917 & 0.00558 & 0.02998 & 0.00817 \\ \cline{2-7}
		& \multirow{4}{*}{0.1}&   50 & 0.22941 & 0.00119 & 0.23764 & 0.00176 \\ \cline{3-7}
		&&   100 & 0.16278 & 0.00210 & 0.16719 & 0.00300 \\ \cline{3-7}
		&&   250 & 0.10096 & 0.00503 & 0.10376 & 0.00702 \\ \cline{3-7}
		&&   500 & 0.06847 & 0.00908 & 0.07014 & 0.01326 \\ \cline{2-7}
		& \multirow{4}{*}{0.5}&   50 & 0.91954 & 0.00339 & 0.94349 & 0.00483 \\ \cline{3-7}
		&&   100 & 0.75020 & 0.00758 & 0.76419 & 0.01086 \\ \cline{3-7}
		&&   250 & 0.56608 & 0.01865 & 0.58265 & 0.02786 \\ \cline{3-7}
		&&   500 & 0.41351 & 0.03322 & 0.42388 & 0.04756 \\ \cline{2-7}
		& \multirow{4}{*}{1}&   50 & 1.15131 & 0.00434 & 1.17831 & 0.00617 \\ \cline{3-7}
		&&   100 & 1.06114 & 0.00926 & 1.08457 & 0.01355 \\ \cline{3-7}
		&&   250 & 0.91941 & 0.02432 & 0.94822 & 0.03475 \\ \cline{3-7}
		&&   500 & 0.76804 & 0.04435 & 0.79063 & 0.06720 \\ \cline{2-7}
		& \multirow{4}{*}{5}&   50 & 1.36798 & 0.00419 & 1.39901 & 0.00582 \\ \cline{3-7}
		&&   100 & 1.30432 & 0.01059 & 1.33375 & 0.01629 \\ \cline{3-7}
		&&   250 & 1.19988 & 0.02431 & 1.22636 & 0.03481 \\ \cline{3-7}
		&&   500 & 1.13442 & 0.04344 & 1.15883 & 0.06197 \\ \cline{2-7}
		& \multirow{4}{*}{10}&   50 & 1.39784 & 0.00435 & 1.43111 & 0.00648 \\ \cline{3-7}
		&&   100 & 1.44322 & 0.01143 & 1.47934 & 0.01659 \\ \cline{3-7}
		&&   250 & 1.37432 & 0.02568 & 1.40912 & 0.03777 \\ \cline{3-7}
		&&   500 & 1.37062 & 0.04267 & 1.40315 & 0.06225 \\ \hline
	\end{longtable}
\end{center}

\begin{center}
	\begin{longtable}{ccccccccccc}
		\caption{
			Average accuracy and elapsed time for the scale parameter estimation example from 100 repeats per setting. Accuracy is measured by the relative error $|\hat{\sigma}_{\text{MLE}}-\sigma_0|/\sigma_0$ and elapsed time is measured in seconds. 		
		} \label{table:extended_concentration}\\
		\hline
		\multirow{2}{*}{$\sigma_0$} & \multirow{2}{*}{$p$} & \multirow{2}{*}{$n$} & \multicolumn{4}{c}{accuracy} & \multicolumn{4}{c}{time} \\ \cline{4-11} 
		&                     &    & \textsf{NewtonE} & \textsf{NewtonA} & \textsf{Roptim} & \textsf{DE} & \textsf{NewtonE} & \textsf{NewtonA} & \textsf{Roptim} & \textsf{DE} \\ \hline
		\multirow{12}{*}{$0.01$}
		& \multirow{4}{*}{5}&50 & 0.08476 & 0.07592 & 0.08391 & 0.08397 & 0.00302 & 0.00286 & 0.00578 & 0.05060 \\ \cline{3-11} 
		&&100 & 0.07152 & 0.06170 & 0.07239 & 0.07233 & 0.00345 & 0.00356 & 0.00774 & 0.04363 \\ \cline{3-11} 
		&&250 & 0.07278 & 0.06250 & 0.07434 & 0.07423 & 0.00603 & 0.00630 & 0.01593 & 0.04498 \\ \cline{3-11} 
		&&500 & 0.07289 & 0.06235 & 0.07437 & 0.07427 & 0.01044 & 0.01047 & 0.02882 & 0.04939 \\ \cline{2-11} 
		& \multirow{4}{*}{10}&50 & 0.18025 & 0.18320 & 0.18324 & 0.21126 & 0.00219 & 0.00270 & 0.00692 & 0.04245 \\ \cline{3-11} 
		&&100 & 0.17433 & 0.17645 & 0.17413 & 0.19400 & 0.00326 & 0.00358 & 0.00921 & 0.04334 \\ \cline{3-11} 
		&&250 & 0.17728 & 0.18015 & 0.17720 & 0.20104 & 0.00566 & 0.00656 & 0.01609 & 0.04625 \\ \cline{3-11} 
		&&500 & 0.17432 & 0.17604 & 0.17460 & 0.19928 & 0.00993 & 0.01045 & 0.03076 & 0.05010 \\ \cline{2-11} 
		& \multirow{4}{*}{20}&50 &  0.33431 &   0.33743 & 0.34380 & 0.34421 & 0.00608 & 0.01412 & 0.00962 & 0.05629 \\ \cline{3-11} 
		&&100 &    0.33646 &    0.33305 & 0.33467 & 0.33383 & 0.00355 & 0.00588 & 0.00977 & 0.04512 \\ \cline{3-11} 
		&&250 &    0.33514 &    0.33203 & 0.33707 & 0.33742 & 0.00176 & 0.00392 & 0.01693 & 0.04721 \\ \cline{3-11} 
		&&500 &    0.33799 &    0.33263 & 0.33543 & 0.33527 & 0.00074 & 0.00279 & 0.03125 & 0.05078 \\ \hline
		\multirow{12}{*}{$0.05$}
		& \multirow{4}{*}{5}&50 & 0.06596 & 0.06595 & 0.06582 & 0.06589 & 0.00244 & 0.00271 & 0.00574 & 0.05216 \\ \cline{3-11} 
		&&100 & 0.04714 & 0.04706 & 0.04710 & 0.04699 & 0.00370 & 0.00424 & 0.00870 & 0.05252 \\ \cline{3-11} 
		&&250 & 0.04827 & 0.04822 & 0.04823 & 0.04828 & 0.00634 & 0.00599 & 0.01692 & 0.05054 \\ \cline{3-11} 
		&&500 & 0.04797 & 0.04800 & 0.04797 & 0.04794 & 0.00980 & 0.01013 & 0.02931 & 0.05446 \\ \cline{2-11} 
		& \multirow{4}{*}{10}&50 & 0.17251 & 0.17250 & 0.17253 & 0.17255 & 0.00211 & 0.00283 & 0.00544 & 0.04097 \\ \cline{3-11} 
		&&100 & 0.15935 & 0.15957 & 0.15945 & 0.15929 & 0.00316 & 0.00333 & 0.00822 & 0.04137 \\ \cline{3-11} 
		&&250 & 0.15437 & 0.15423 & 0.15414 & 0.15409 & 0.00555 & 0.00549 & 0.03549 & 0.04221 \\ \cline{3-11} 
		&&500 & 0.15557 & 0.15546 & 0.15563 & 0.15567 & 0.00931 & 0.00957 & 0.02636 & 0.04556 \\ \cline{2-11} 
		& \multirow{4}{*}{20}&50 &    0.30124 &    0.30467 & 0.30743 & 0.30756 & 0.00493 & 0.00406 & 0.00617 & 0.03738 \\ \cline{3-11} 
		&&100 &    0.29983 &  0.29687 & 0.30345 & 0.30363 & 0.00521 & 0.00608 & 0.00830 & 0.04097 \\ \cline{3-11} 
		&&250 &    0.29234 &  0.29548 & 0.29978 & 0.29932 & 0.00828 & 0.00852 & 0.01636 & 0.04442 \\ \cline{3-11} 
		&&500 &    0.29043 &  0.29635 & 0.29891 & 0.29896 & 0.01104 & 0.01163 & 0.03167 & 0.04748 \\ \hline 
		\multirow{12}{*}{$0.1$}
		& \multirow{4}{*}{5}&50 & 0.08211 & 0.08203 & 0.08227 & 0.08214 & 0.00211 & 0.00256 & 0.00599 & 0.04470 \\ \cline{3-11}
		&&100 & 0.05459 & 0.05471 & 0.05462 & 0.05460 & 0.00297 & 0.00344 & 0.00975 & 0.04411 \\ \cline{3-11}
		&&250 & 0.04692 & 0.04697 & 0.04688 & 0.04689 & 0.00599 & 0.00588 & 0.01668 & 0.04658 \\ \cline{3-11}
		&&500 & 0.04376 & 0.04371 & 0.04369 & 0.04371 & 0.00954 & 0.01075 & 0.02782 & 0.04950 \\ \cline{2-11}
		& \multirow{4}{*}{10}&50& 0.12741 & 0.12717 & 0.12744 & 0.12733 & 0.00239 & 0.00257 & 0.00550 & 0.04220 \\ \cline{3-11}
		&&100& 0.12658 & 0.12642 & 0.12655 & 0.12654 & 0.00308 & 0.00377 & 0.00842 & 0.04436 \\ \cline{3-11}
		&&250& 0.11350 & 0.11333 & 0.11333 & 0.11311 & 0.00599 & 0.00691 & 0.01566 & 0.04757 \\ \cline{3-11}
		&&500& 0.11697 & 0.11693 & 0.11705 & 0.11689 & 0.01026 & 0.01068 & 0.02884 & 0.05080 \\ \cline{2-11}
		& \multirow{4}{*}{20}&50& 0.15965 & 0.15970 & 0.15980 & 0.15966 & 0.00253 & 0.00264 & 0.00615 & 0.04245 \\ \cline{3-11}
		&&100& 0.14674 & 0.14684 & 0.14691 & 0.14704 & 0.00308 & 0.00342 & 0.00947 & 0.04818 \\ \cline{3-11}
		&&250& 0.14730 & 0.14706 & 0.14715 & 0.14731 & 0.00595 & 0.00640 & 0.01826 & 0.04716 \\ \cline{3-11}
		&&500& 0.14029 & 0.14028 & 0.14010 & 0.14016 & 0.01050 & 0.01128 & 0.03038 & 0.05181 \\ \hline
		\multirow{12}{*}{$0.5$}
		& \multirow{4}{*}{5}&50 & 0.13775 & 0.13774 & 0.13769 & 0.13761 & 0.00207 & 0.00267 & 0.00547 & 0.03804 \\ \cline{3-11}
		&&100 & 0.09126 & 0.09133 & 0.09134 & 0.09146 & 0.00311 & 0.00383 & 0.00888 & 0.03870 \\ \cline{3-11}
		&&250 & 0.06190 & 0.06173 & 0.06183 & 0.06192 & 0.00567 & 0.00632 & 0.01571 & 0.04117 \\ \cline{3-11}
		&&500 & 0.03962 & 0.03956 & 0.03953 & 0.03954 & 0.01002 & 0.01068 & 0.03084 & 0.05024 \\ \cline{2-11}
		& \multirow{4}{*}{10}&50& 0.21176 & 0.21193 & 0.21203 & 0.21195 & 0.00234 & 0.00258 & 0.00557 & 0.04184 \\ \cline{3-11}
		&&100& 0.13793 & 0.13808 & 0.13804 & 0.13784 & 0.00319 & 0.00346 & 0.00848 & 0.04343 \\ \cline{3-11}
		&&250& 0.10370 & 0.10404 & 0.10392 & 0.10372 & 0.00541 & 0.00653 & 0.01501 & 0.04461 \\ \cline{3-11}
		&&500& 0.06312 & 0.06310 & 0.06319 & 0.06321 & 0.01013 & 0.01039 & 0.02857 & 0.04812 \\ \cline{2-11}
		& \multirow{4}{*}{20}&50& 0.42899 & 0.42865 & 0.42862 & 0.42841 & 0.00254 & 0.00262 & 0.00613 & 0.04667 \\ \cline{3-11}
		&&100& 0.28234 & 0.28228 & 0.28239 & 0.28221 & 0.00337 & 0.00378 & 0.00820 & 0.04645 \\ \cline{3-11}
		&&250& 0.19248 & 0.19208 & 0.19251 & 0.19237 & 0.00555 & 0.00595 & 0.01591 & 0.04761 \\ \cline{3-11}
		&&500& 0.11190 & 0.11188 & 0.11161 & 0.11188 & 0.00960 & 0.01537 & 0.02721 & 0.04991 \\ \hline
		\multirow{12}{*}{$1$}
		& \multirow{4}{*}{5}&50 & 0.22924 & 0.22915 & 0.22935 & 0.22958 & 0.00201 & 0.00266 & 0.00591 & 0.03786 \\ \cline{3-11}
		&&100 & 0.16745 & 0.16695 & 0.16710 & 0.16734 & 0.00277 & 0.00341 & 0.00870 & 0.03744 \\ \cline{3-11}
		&&250 & 0.12082 & 0.12102 & 0.12087 & 0.12087 & 0.00540 & 0.00613 & 0.01574 & 0.03923 \\ \cline{3-11}
		&&500 & 0.09411 & 0.09406 & 0.09396 & 0.09399 & 0.01006 & 0.01030 & 0.02743 & 0.04383 \\ \cline{2-11}
		& \multirow{4}{*}{10}&50& 0.44890 & 0.44828 & 0.44850 & 0.44842 & 0.00207 & 0.00308 & 0.00567 & 0.04217 \\ \cline{3-11}
		&&100& 0.29815 & 0.29752 & 0.29777 & 0.29775 & 0.00302 & 0.00350 & 0.00930 & 0.04312 \\ \cline{3-11}
		&&250& 0.19481 & 0.19477 & 0.19517 & 0.19503 & 0.00578 & 0.00640 & 0.01586 & 0.04965 \\ \cline{3-11}
		&&500& 0.15321 & 0.15316 & 0.15329 & 0.15339 & 0.00941 & 0.00988 & 0.02685 & 0.04639 \\ \cline{2-11}
		& \multirow{4}{*}{20}&50& 0.65159 & 0.65101 & 0.65132 & 0.65158 & 0.00253 & 0.00288 & 0.00625 & 0.04664 \\ \cline{3-11}
		&&100& 0.53600 & 0.53655 & 0.53708 & 0.53731 & 0.00310 & 0.00371 & 0.00841 & 0.04766 \\ \cline{3-11}
		&&250& 0.33806 & 0.33847 & 0.33757 & 0.33878 & 0.00591 & 0.00575 & 0.01612 & 0.04525 \\ \cline{3-11}
		&&500& 0.25429 & 0.25480 & 0.25491 & 0.25456 & 0.01012 & 0.00956 & 0.02610 & 0.04889 \\ \hline
		\multirow{12}{*}{$5$}
		& \multirow{4}{*}{5}&50 & 0.71740 & 0.71967 & 0.71801 & 0.71752 & 0.00212 & 0.00294 & 0.00662 & 0.04423 \\ \cline{3-11}
		&&100 & 0.63575 & 0.63764 & 0.63610 & 0.63647 & 0.00314 & 0.00399 & 0.00765 & 0.03694 \\ \cline{3-11}
		&&250 & 0.45104 & 0.45045 & 0.45059 & 0.45139 & 0.00573 & 0.00666 & 0.01527 & 0.03954 \\ \cline{3-11}
		&&500 & 0.35416 & 0.35469 & 0.35444 & 0.35491 & 0.01027 & 0.01108 & 0.02760 & 0.04411 \\ \cline{2-11} 
		& \multirow{4}{*}{10}&50& 0.85163 & 0.85407 & 0.85222 & 0.85366 & 0.00254 & 0.00310 & 0.00600 & 0.04527 \\ \cline{3-11}
		&&100& 0.80135 & 0.80051 & 0.80126 & 0.80144 & 0.00365 & 0.00376 & 0.00997 & 0.04395 \\ \cline{3-11}
		&&250& 0.67648 & 0.67539 & 0.67657 & 0.67521 & 0.00605 & 0.00709 & 0.01707 & 0.04646 \\ \cline{3-11}
		&&500& 0.51264 & 0.51321 & 0.51206 & 0.51318 & 0.01171 & 0.01271 & 0.03043 & 0.05196 \\ \cline{2-11}
		& \multirow{4}{*}{20}&50& 0.93176 & 0.93154 & 0.93079 & 0.93204 & 0.00238 & 0.00348 & 0.00687 & 0.04793 \\ \cline{3-11}
		&&100& 0.90124 & 0.90096 & 0.89997 & 0.90214 & 0.00398 & 0.00370 & 0.00940 & 0.05832 \\ \cline{3-11}
		&&250& 0.84475 & 0.84479 & 0.84456 & 0.84534 & 0.00609 & 0.00663 & 0.01450 & 0.04643 \\ \cline{3-11}
		&&500& 0.78437 & 0.78493 & 0.78330 & 0.78365 & 0.01013 & 0.01049 & 0.02714 & 0.05347 \\ \hline
		\multirow{12}{*}{$10$}
		& \multirow{4}{*}{5}&50 & 0.85464 & 0.85690 & 0.85581 & 0.85660 & 0.00218 & 0.00363 & 0.00543 & 0.03934 \\ \cline{3-11}
		&&100 & 0.81485 & 0.81529 & 0.81429 & 0.81520 & 0.00334 & 0.00429 & 0.00816 & 0.03856 \\ \cline{3-11}
		&&250 & 0.70522 & 0.70388 & 0.70391 & 0.70506 & 0.00613 & 0.00733 & 0.01706 & 0.04160 \\ \cline{3-11}
		&&500 & 0.65070 & 0.64979 & 0.65008 & 0.65012 & 0.01006 & 0.01129 & 0.03010 & 0.04922 \\ \cline{2-11}
		& \multirow{4}{*}{10}&50& 0.92874 & 0.92971 & 0.93049 & 0.93003 & 0.00245 & 0.00264 & 0.00582 & 0.04183 \\ \cline{3-11}
		&&100& 0.90104 & 0.90136 & 0.90001 & 0.90096 & 0.00348 & 0.00405 & 0.00821 & 0.04234 \\ \cline{3-11}
		&&250& 0.84038 & 0.84082 & 0.84018 & 0.84047 & 0.00618 & 0.00677 & 0.01516 & 0.04610 \\ \cline{3-11}
		&&500& 0.78663 & 0.78595 & 0.78569 & 0.78597 & 0.01078 & 0.01127 & 0.03031 & 0.05010 \\ \cline{2-11}
		& \multirow{4}{*}{20}&50& 0.96752 & 0.96667 & 0.96787 & 0.96733 & 0.00260 & 0.00281 & 0.00842 & 0.05209 \\ \cline{3-11}
		&&100& 0.94684 & 0.94531 & 0.94769 & 0.94665 & 0.00334 & 0.00345 & 0.00868 & 0.04889 \\ \cline{3-11}
		&&250& 0.91812 & 0.92019 & 0.91929 & 0.91853 & 0.00610 & 0.00553 & 0.01623 & 0.04735 \\ \cline{3-11}
		&&500& 0.88335 & 0.88276 & 0.88389 & 0.88446 & 0.01002 & 0.01043 & 0.02840 & 0.05380 \\ \hline
	\end{longtable}
\end{center}
\end{document}